\documentclass[fleqn,usenatbib]{mnras}

\usepackage{newtxtext,newtxmath}

\usepackage[T1]{fontenc}

\DeclareRobustCommand{\VAN}[3]{#2}
\let\VANthebibliography\thebibliography
\def\thebibliography{\DeclareRobustCommand{\VAN}[3]{##3}\VANthebibliography}

\usepackage{graphicx}	
\usepackage{amsmath}	

\title{Migration and Mixing in the Galactic Disc from Encounters between Sagittarius and the Milky Way}

\author[Carr et al.]{
Christopher Carr,$^{1}$\thanks{E-mail: cc4504@columbia.edu}
Kathryn V. Johnston,$^{1,2}$
Chervin F. P. Laporte$^{3}$
and Melissa K. Ness$^{1,2}$
\\
$^{1}$Department of Astronomy, Columbia University, 550 West 120th Street, New York, NY, 10027, U.S.\\
$^{2}$Center for Computational Astrophysics, Flatiron Institute, 162 5th Avenue, Manhattan, NY, USA\\
$^{3}$Institut de Ciències del Cosmos (ICCUB), Universitat de Barcelona (IEEC-UB), Martí i Franquès 1, E-08028 Barcelona, Spain
}

\date{Accepted XXX. Received YYY; in original form ZZZ}

\pubyear{2015}

\begin{document}
\label{firstpage}
\pagerange{\pageref{firstpage}--\pageref{lastpage}}
\maketitle

\begin{abstract}
Stars born on near-circular orbits in spiral galaxies can subsequently migrate to different orbits due to interactions with non-axisymmetric disturbances within the disc such as bars or spiral arms. This paper extends the study of migration to examine the role of external influences using the example of the interaction of the Sagittarius dwarf galaxy (Sgr) with the Milky Way (MW). We first make impulse approximation estimates to characterize the influence of Sgr disc passages. The tidal forcing from Sgr can produce changes in both guiding radius $\Delta R_{g}$  and orbital eccentricity, as quantified by the maximum radial excursion $\Delta R_{\rm max}$. These changes follow a quadrupole-like pattern across the face of the disc, with amplitude increasing with Galactocentric radius. We next examine a collisionless N-body simulation of a Sgr-like satellite interacting with a MW-like galaxy and find that Sgr’s influence in the outer disc dominates over the secular evolution of orbits between disc passages. Finally, we use the same simulation to explore possible observable signatures of Sgr-induced migration by painting the simulation with different age stellar populations. We find that following Sgr disc passages, the migration it induces manifests within an annulus as  an approximate quadrupole in azimuthal metallicity variations ($\delta_{\rm [Fe/H]}$), along with systematic variations in orbital eccentricity, $\Delta R_{\rm max}$. These systematic variations can persist for several rotational periods. We conclude that this combination of signatures may be used to distinguish between the different migration mechanisms shaping the chemical abundance patterns of the Milky Way's thin disc.  
\end{abstract}




\section{Introduction}
With the Gaia-Enceladus-Sausage merger believed to be at least $8-11$ Gyr ago \citep{2018_Helmi_merger, 2018_belokurov_co-formation}, the thin disc is thought to have grown smoothly from the inside-out, first accreting cool gas before forming stars that chemically enrich the interstellar medium. Such gaseous discs birth stars on nearly circular orbits with tight correlative properties, such as age and abundance gradients that decrease with increasing Galactocentric radius  \citep[e.g.][]{2019_Frankel_inside-out}. If this were the whole story, then the present-day orbits of stars and the disc's global chemical properties ought to reflect closely the conditions of their formation, but disc structure and perturbations unsettle this simple narrative. 

The thin disc is an active site of interest for investigating the dynamical drivers of the Galaxy's evolution \citep{freeman_new_2002, 2014_sellwood_secular}. Interactions with non-asymmetric disc features$-$spiral arms, bars, and molecular clouds$-$have been shown to significantly alter the properties of stellar orbits, redistributing stars onto new orbits that place them at Galactocentric radii radically different from their birth radii \citep{sellwood_radial_2002, minchev_estimating_2018}. This process of altering the properties of stellar orbits through the gravitational encounter with a non-axisymmetric potential is broadly defined as radial migration. There has been much attention paid to the secular drivers of this phenomena \citep{sellwood_radial_2002, 2008_roskar_spiral_waves, roskar_radial_2012,minchev_new_2010,kubryk_radial_2013}. Radial migration has taken on many different meanings in the literature since its introduction in \citet{sellwood_radial_2002}. It was originally intended to describe the change in angular momentum of a stellar orbit through a resonant interaction with a spiral arm. Radial migration of this kind, including other dynamical interactions that can alter the angular momentum of an orbit without adding any excess random energy, has come to be referred to as \textit{churning}. Dynamical processes that heat an orbit without an increase in angular momentum is called \textit{blurring}. These secular processes have been shown to be quite significant in disc simulations, causing migration on a scale of several kpc on Gyr timescales \citep{roskar_radial_2012}. 

Migrated and non-migrated populations do not leave much to distinguish themselves by their orbital properties. However, the mixing of different stellar populations does leave a discernible mark on the age and chemical properties of the galaxy. The chemistry of stars, namely their [Fe/H] abundance, is a measure of their birth environment and conserved over a stellar lifetime. Radial migration has been invoked to explain the large scatter in the Age-Metalicity Relation (AMR) in the solar neighborhood \citep{sellwood_radial_2002, 2007_Binney_dynamics}, the observed change in skewness of the metallicity distribution function as a function of Galactocentric radius \citep{2015_Hayden_APOGEE, loebman_imprints_2016, 2016_Martinez-Mediana_MDF}, and the flattening of age and metallicity gradients \citep{2005A&A_Maciel_gradients, 2015A&A_kubryk_abundance-profiles, 2020_Vincenzo_metal-flows}. There has also been work on the impact of radial migration, brought about from spiral structure or a galactic bar, on azimuthal variations in the metallicity distribution \citep{di_matteo_signatures_2013, 2016_grand_spirals-patterns, 2021_wheeler}.

The broad recognition of radial migration as a significant mechanism that can reshape the dynamical and chemical composition of the galaxy has motivated some to construct models that can unravel its effects. By assuming a certain time evolution of the metallicity gradient and diffusion timescale for radial migration, we can build models that return stars to their ``birth radii" and recover the Galaxy's formation properties \citep{minchev_estimating_2018, 2018-Frankel_orbit-migration, 2019_Frankel_inside-out, Frankel_cool}. These efforts have shown some promise in illuminating the properties of extended secular modes of radial migration under the assumption of an isolated galaxy. However, the history of the Galaxy is under no obligation to be this simple. 

External influences on the disc, such as satellite bombardment, can also drive large-scale migration across galaxies. For the case of the Milky Way, the ongoing disruption and merger with the Sagittarius dwarf spheroidal galaxy (Sgr) points to an important interruption in the Galaxy’s relatively quiescent history. Since its discovery \citep{ibata_dwarf_1994}, its historical interaction with the Milky Way has been cited as a possible mechanism for the myriad disequilibrium structures. These include  
vertical warp modes in the gas and stellar discs \citep{1998A_Ibata_HI_warps, 2003_Bailin_coupling}, dynamically cold rings like the Monoceros Ring \citep{younger_origin_2008}, spiral arms \citep{purcell_sagittarius_2011}, and the ``phase-spiral" observed in the $z-v_z$ phase plane \citep{2018_antoja_snail}. Despite these and recent efforts to deconstruct the contribution of Sgr on the vertial response of the disc \citep{2013_gomez_vertical_density_waves, laporte_response_2018, 2021_Poggio_vertical_response}, its full influence on radial migration has not been fully considered. 

The role of satellite perturbers in the radial redistribution of stars has been explored in past studies. Using a test particle disc simulation and a $10^9 M_{\odot}$ satellite on a Sgr-like orbit, \citet{quillen_radial_2009} discovered a distinct population of low-eccentricity stars with origins in the outer disc migrating into the solar neighborhood. Galactic discs situated within a $\Lambda$CDM-motivated cosmology undergo significant radial mixing and can exhibit qualities that distinguish it from non-perturbed discs, such as an increased fraction of migrating stars into the inner regions of the galaxy from the outer disc \citep{bird_radial_2012} In addition, enhanced migration from early major mergers has been invoked to explain the population of old $\alpha$-enhanced stars with low velocity dispersion in the solar neighborhood \citep{2013_minchev_sol_vicinity, 2014_minchev_relations}, a scenario consistent with \textit{Gaia}-ESO survey observations \citep{2015_guiglion_constraints, 2018_hayden_churning}. More recent work from \citet{2021_turning-points_Lu} reveals that an encounter with a low-mass satellite and the subsequent radial migration may be the main culprit behind the observed turning points in the Milky Way's AMR \citep{2019_Feuillet_spatial_variations}. It is clear that satellite encounters can drive radial migration on a scale comparable to that observed from secular mechanisms. However, there is still more work to be done if we are to gain a better grasp of what the Milky Way's unique history of satellite encounters has had on the properties of its stellar orbits. Furthermore, to determine if there are any persisting signatures today that could properly illuminate this history. 

Here we seek to build on the established literature on radial migration by focusing on angular momentum evolution and radial heating of orbits in a simulation. The simulation that we use for this is an analogue of the Milky Way's ongoing interaction with the Sagittarius dwarf galaxy. This allows us to study in detail how orbits respond to the disc crossings of Sgr. We can also test to what degree the overall $J_\phi$-evolution of stellar orbits over the course of the simulation can be attributed to the short-term forcing from Sgr, as opposed to the extended, secular processes internal to the disc. After detailing the properties of the simulation and our methods in section \ref{sec: 2}, we describe the evolution of Sagittarius' orbital trajectory and mass in section \ref{sec: 3}. In section \ref{sec: 4}, we introduce the impulse approximation to get a qualitative scale of the disc's response to Sgr, we follow the extended orbit evolution between encounters from spiral arms, and finally, contrast the two modes of migration \& mixing. In section \ref{sec: 5}, we discuss potential observational traces from the most recent Sgr disc crossings in the metallicity distribution of stars in the solar annulus and outer disc. Finally, we then discuss the implications of our results for studies of radial migration in the Milky Way and future prospects for our work in section \ref{sec: 6}, and summarize our results in section \ref{sec: 7}. 

\begin{figure}
	\centering
  	\includegraphics[width=1\linewidth]{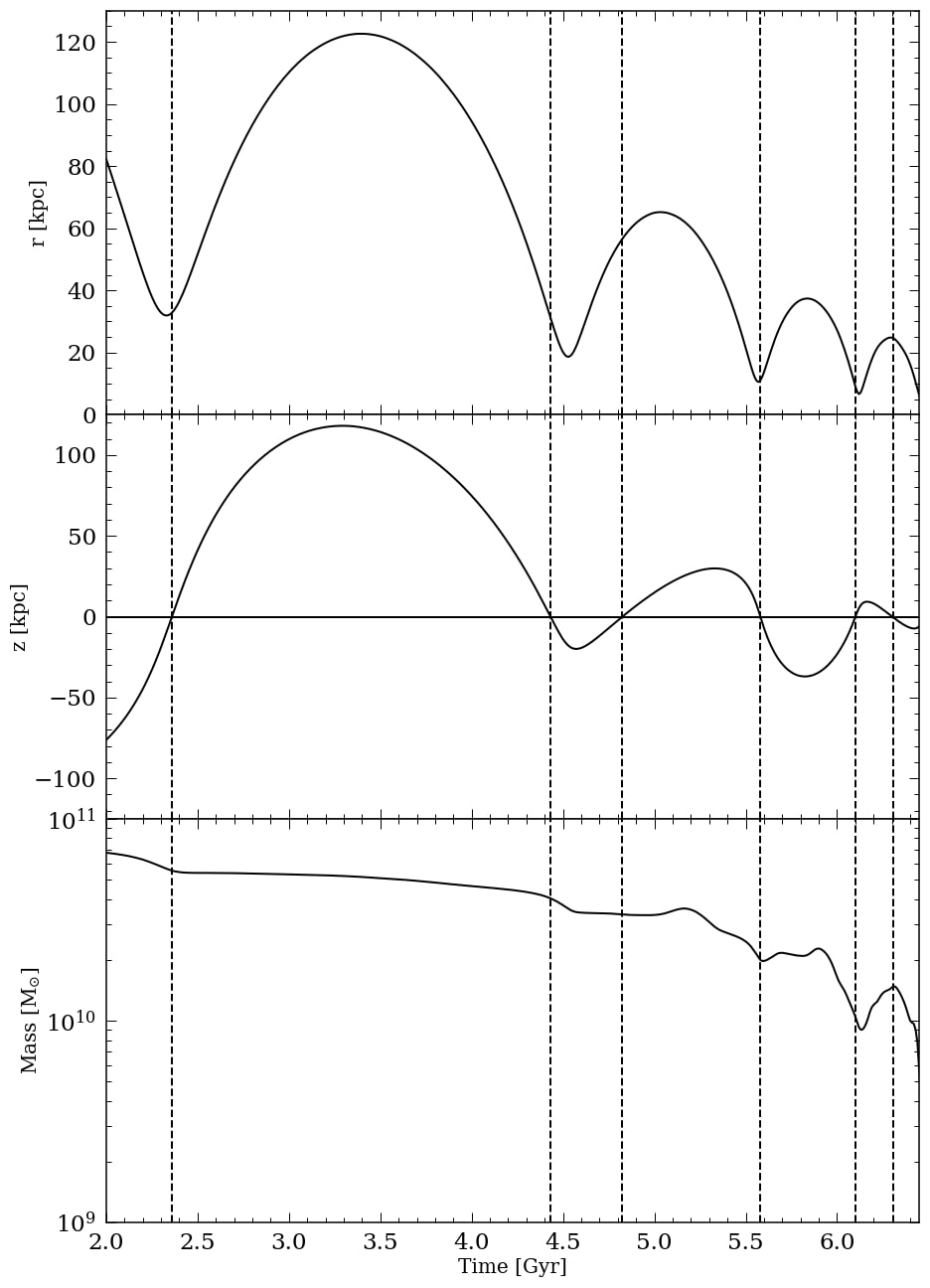}
    \caption{Orbit of Sagittarius charted by the motion of its center of mass in Galactocentric radius (\textbf{top}) and height above the disc (\textbf{middle}), and the bounded mass of Sagittarius over the course of the simulation (\textbf{bottom}). Dashed vertical lines in all plots denote the pericenter crossings of the $z=0$ midplane.}
    \label{fig: orbit}
\end{figure}

\begin{figure*}
	\centering
  	\includegraphics[width=1\textwidth, scale=1]{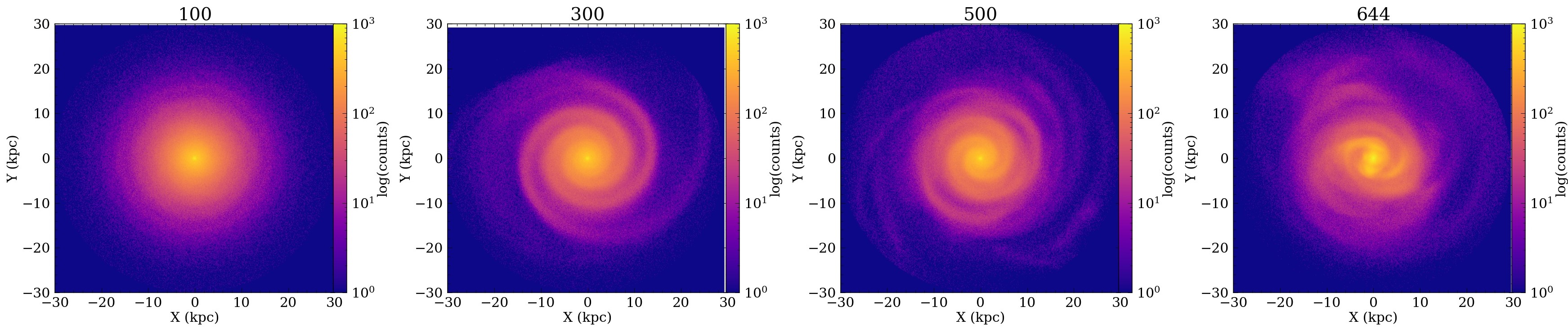}
    \caption{Facedown projections of particle density during different regimes of the simulation. Snapshots are taken at times $t=1.0$ Gyr, $t=3.0$ Gyr, $t=5.0$ Gyr, and $t=6.44$ Gyr respectively. Snapshot $t=6.44$ Gyr is our assumed present-day snapshot.}
    \label{fig: density}
\end{figure*}

\section{Numerical Methods} \label{sec: 2}

In this section, we describe the simulation used for our analysis, followed by a brief introduction of the action coordinates and our criteria for selecting particles on circular orbits. 

\subsection{Simulation Description}

For this work, we use the simulations of \citet{laporte_influence_2018}. These are a set of collisionless N-body simulations that consider the interaction between Sgr and the Galactic disc, including realizations that combine the effects of Sgr and the Large Magellanic Cloud (LMC) on its first infall orbit. We focus on the L2 model, which deals with the Galactic disc and Sgr in isolation. Other more massive progenitor models produced vertical oscillations in the disc with amplitudes in excess to those seen in the L2 model and are not presented here for clarity \citep{laporte_response_2018}. 

 We now give a brief description of the simulation's parameters, and we refer to the relevant contributions for a more thorough discussion \citep{laporte_influence_2018}. In these runs, the host MW has the following properties: dark halo of $M_h$ = 10$^{12} M_{\odot}$, an exponential disc of $M_{disc}$ = $6 \times 10^{10} M_{\odot}$, with a scale length of $R_d$ = 3.5 kpc, scale height of $h_d$ = 0.53 kpc and a central bulge with mass of $M_{bulge} = 10^{10} M_{\odot}$ and scale radius, $a_b = 0.7$ kpc. This choice of parameters results in a circular velocity $V_{circ}$ = 239 km/s at $R_0$ = 8 kpc at the end of simulation. The Toomre Q parameter is $>$ 1 everywhere in the disc to maintain disc stability over the course of the simulation, minimizing the spontaneous formation of disc structure with origins separate from those induced by Sgr.
 
Sgr has a starting virial mass of $M_{sgr} = 8 \times 10^{10} M_{\odot}$ at the start of the simulation. Its \citet{hernquist_analytical_1990} profile has the corresponding parameterization of $M_h = 8 \times 10^{10} M_{\odot}$ and scale length, $a_h = 8$ kpc. Sgr also includes a subdominant Hernquist profile of $M_{*} = 6.4 \times 10^{8}$ and $a_{*} = 0.85$ kpc to represent the stellar component of the dwarf galaxy. Particle masses of $m_{h,sgr} = 2 \times 10^{4} M_{\odot}$ and $m_{*, sgr} = 4 \times 10^{3}$ were used to represent the dark matter and stellar component respectively. 

The N-body simulation is ran with the tree-code GADGET$-$3 code \citep{springel_cosmological_2005}. Particle masses of $m_h$ = 2.6 $\times 10^4 M_{\odot}$, $m_d$ = $1.2 \times 10^4 M_{\odot}$, and $m_b$ = $10^4 M_{\odot}$ are used to represent the dark matter halo, disc and bulge components of the Galaxy respectively. We use softening lengths for the halo of $\epsilon_h$ = 60pc, and equivalent softening lengths for the disc and bulge, $\epsilon_d$ = $\epsilon_b$ = 30pc. We represent the disc with $N_d \sim 5 \times 10^6$ particles and the halo with $N_h \sim 4 \times 10^7$ particles. 

\subsection{Action Calculation}
We calculate actions using particle positions and velocities that are centered with respect to the galactic bulge, and aligned the rotational axis of the disc to the z-axis of the rotation matrix for all snapshots in the simulation.  
Action-angle variables for particles in the disc were calculated using AGAMA, a software library for broad application in stellar dynamics and galaxy-modeling \citep{vasiliev_agama_2019}. AGAMA generates a smooth approximation of the potential, and estimates actions in the disc using the St\"{a}ckel fudge method. 

In this paper we use the three standard action coordinates, which are conserved quantities that describe an orbit in an axisymmetric potential. We compute the vertical action ($J_z$), the radial action ($J_R$), and the azimuthal action ($J_\phi$), where $J_z$ and $J_R$ describe the oscillations of the orbit in the vertical and radial directions respectively. In an axisymmetric potential, $J_\phi$ is equivalent to the z-component of the angular momentum $L_z = R V_{\phi}$, so for the remainder of this work, we will use $J_\phi$ to represent angular momentum. 

\subsection{Selecting circular orbits to characterize Zero-Age Populations}
In order to understand how stars in the disc respond to Sgr, we first need to capture these orbits.  We assume that new stars are born on roughly circular orbits in the thin disc, and we develop a criteria for selecting these orbits in the simulation. We select particles on circular orbits  at time $T_0$ in the simulation and examine the orbital properties of this "stellar population" at subsequent time $T_1$ when it has an ``age" of $T_1-T_0$, in order to follow the dynamical evolution. Note that stars may find themselves on circular orbits early in the simulation when the disc is unperturbed, but also at late times, when particles settle into new orbits following close encounters from Sgr, hence giving us particles that represent both "old" and "young" populations at any given time. Tracking the evolution of these orbits over time allows us to characterize the dynamical imprints of different dynamical processes, and more easily identify significant changes in their energy and angular momentum. We define our circular orbit criteria, $\xi$, by comparing the ratio of their azimuthal actions to the sum in quadrature of their radial and vertical action coordinates,
\begin{equation}
    \xi = \frac{\sqrt{J_R^2 + J_z^2}}{J_{\phi}}.
\end{equation}
We set the threshold for circularity at $\xi = 2.5 \times 10^{-3}$. This threshold was set by the lowest 5$\%$ of orbits with respect to our criteria for the first simulation snapshot, and we maintain a consistent threshold of $\xi = 2.5 \times 10^{-3}$ throughout our work. This choice of threshold caps radial and vertical oscillations just under $6 \%$ of the guiding center radius, so for an orbit with $R_g = 8$ kpc, this would correspond to a maximum oscillation of $\sim 0.5$ kpc in the radial or vertical directions to be considered a near-circular orbit. Radial velocities of these circular orbits are normally distributed around a mean of $\sim 0$ km/s with a standard deviation $\sim 8$ km/s.

\section{Satellite Evolution} \label{sec: 3}

\subsection{Orbital Properties of Sagittarius and Characterizing Evolutionary Regimes of the Disc}
The orbit of the satellite was placed as close as possible to a Sgr-like orbit in order to best model the influence on Sgr on the structure of the disc. Detailed fully in \citep{laporte_influence_2018}, the Sgr-like orbit is constrained by the present-day location, shape, and line-of-sight velocity of the Sgr stream \citep{majewski_two_2003}. With the implementation of dynamical friction \citep{chandrasekhar_dynamical_1943}, the orbit is backward integrated from the initial conditions of the Sgr stream's current properties to find Sgr's position and velocity vector at edge of the MW's virial radius ($R_{200}\sim 214$ kpc). 

Figure \ref{fig: orbit} displays the resulting orbital decay of Sgr for the L2 model, with vertical lines marking each passage of the $Z=0$ plane. The time and length scale of the satellite's orbit provide a framework for characterizing regimes of disc bombardment. Figure \ref{fig: density} displays how this bombardment affects disc structure through the facedown projections in particle density. The first period of the simulation--approximately lasting 2 Gyr--capture the disc in an era of isolation before the first pericenter passage of Sgr. The next period from 2-4 Gyrs, follow the evolution of the disc after the impact of the disc at $t \sim 2.35$ Gyr, and the ensuing response on the particle orbits. The proceeding period between 4-6 Gyr marks an era of infrequent bombardment by the orbiting Satellite at impacts of decreasing Galactocentric radius, while the final regime, 6-7 Gyr, characterize the response of the disc to several repeated passages as Sgr undergoes severe disruption. Prior work from \citet{laporte_influence_2018} has found a qualitative match between stream morphology and line of sight velocities with those observed in \citet{majewski_two_2003} at $t \sim 6.44$ Gyr in simulation time. Therefore, for the purposes of our work, we consider that to be our ''present-day" snapshot for which we will base our analysis. 

\subsection{Mass Loss of Sagittarius Over Time}
Tidal stripping from the disc as Sgr spirals inward causes Sgr to lose mass as it proceeds along its orbit. In Figure \ref{fig: orbit}, we track the evolution of Sgr's bound mass throughout the course of the simulation. This was done by constructing a smooth approximation of the potential using AGAMA's built-in multipole expansion for spherical-like potentials. Particles with kinetic energies less than the value of the approximated potential at their positions are considered bound to the satellite. The bound mass of Sgr begins at $\sim 8 \times 10^{10} M_{\odot}$ and gradually decreases through the run of the simulation until reaching a present day total mass of $\sim 6 \times 10^{9} M_{\odot}$ (see discussion for limitations of this model). Sgr experiences great disruption in the snapshots immediately surrounding the present-day snapshot, which will introduce uncertainty in our algorithm for estimating the satellite potential and mass in this regime. 
 \begin{figure*}
	\centering
  	\includegraphics[width=1\textwidth]{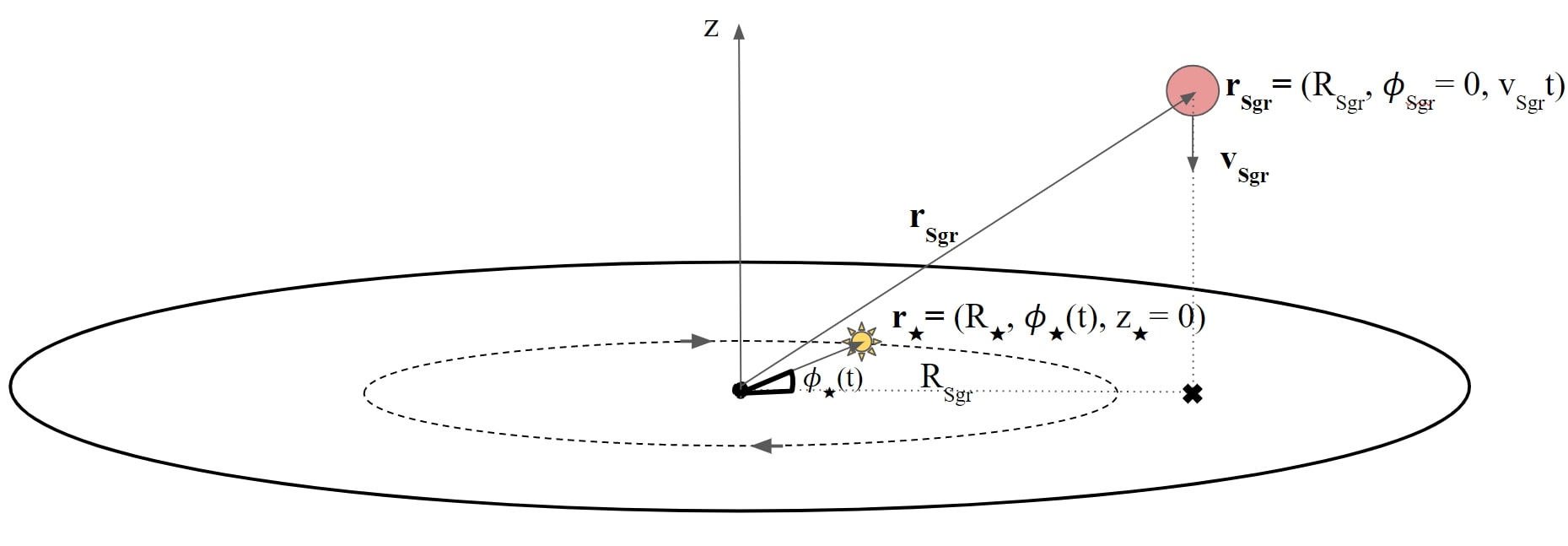}
    \caption{A schematic of the Galactic disc, a star with the position vector $\textbf{r}_{\star}$ with Galactocentric radius $R_{*}$, and Sagittarius with position vector $\textbf{r}_{sgr}$. Sagittarius' orbital trajectory is largely perpendicular to the plane of the disc. (Fig credit: Suroor Seher Gandhi)}
    \label{fig: scheme}
\end{figure*}
\section{Disc Evolution} \label{sec: 4}

In this section we separate and quantify the influence of Sgr on migration and radial mixing for particles initially on circular orbits. Here we use the term radial migration to refer to a change in guiding center of an orbit $\Delta R_g$, which is equivalent to a change in angular momentum (i.e. azimuthal action), $\Delta J_\phi$. We use the term \textit{radial mixing} to refer to the radial heating of an orbit, which marks itself as a change in radial velocity, $\Delta v_R$, leading to a change in the radial action, $\Delta J_R$ and an amplification of the maximum radial excurison, $\Delta R_{\rm max}$, the distance between apocentre and pericentre of an orbit. 

Actions are typically viewed as a conserved quantity of an orbit, and this is true for axisymmetric time-independent potentials or for potentials where the variation of the potential is gradual enough such that stars at different phases of the orbit experience the same time-averaged effect. However, changes to the potential that are neither gradual nor non-axisymmstric, such as an agitation from a nearby satellite or a resonant interaction with a spiral arm, can alter the properties of orbits, changing their action coordinates as well. The change in action coordinates for individual orbits and for stars across the disc through the course of the simulation offers a chance to separate and quantify radial migration and mixing as a consequence of the disc's repeated encounters with Sagittarius (Section \ref{sec: 4.1}), and the secular migration processes that rule the galaxy in the intervening eras between encounters (Section \ref{sec: 4.2}). The two are compared in Section \ref{sec: 4.3}.

\subsection{Response of the Disc to Close Encounters with Sagittarius} \label{sec: 4.1}

In order to clearly separate the influence of Sgr from secular processes, we first examine to what extent we can characterize and understand its contribution using simple tools. To estimate the effect an encounter with Sgr could have on orbits in the disc, we treat the first disc crossing of Sgr as an impulsive encounter. By ``impulsive", we mean that the timescale of the encounter, $\tau_{\rm enc}$ is short with respect to the orbital time, $\tau_{\rm orb}$, for particles in the disc. The impulse approximation is typically evoked to study high-speed encounters, where $\tau_{\rm enc} \ll \tau_{\rm cross}$, however it has been shown to produce accurate results for encounter times on the same order as the crossing time \citep{aguilar_tidal_1985}. In addition, an impulsive treatment of the encounter with Sgr has been used to model the vertical phase-spiral observed in Gaia DR2 \citep{binney_schonrich_2018}, and \citet{bland-hawthorn_impulse2021} has recently explored simplified N-body models of Sgr as a point mass encounter. We demonstrate in the following section that the impulse approximation has remarkable utility in characterizing the response of the disc during the course of the first encounter with Sagittarius and can inform our understanding on the degree of radial migration \& mixing. 

\subsubsection{Estimates for a Single Impact Using the Impulse Approximation} 
Referring to the schematic in Figure \ref{fig: scheme}, the tidal acceleration experienced by each particle in the disc$-$the residual acceleration that remains after subtracting the acceleration due to Sgr at the Galactic center$-$is expressed as 
\begin{equation}
    \textbf{a}_t (t) = -\frac{G M_{\rm sgr}\textbf{r}}{r^3} + \frac{G M_{\rm sgr}\textbf{r}_{\rm sgr}}{r_{\rm sgr}^3},
\end{equation}
where \textbf{r} is the position vector of a particle in the disc with respect to Sgr, while $\textbf{r}_{\rm sgr}$ is the vector between the Galactic center and Sgr, which we treat as a point mass. In the event of a high-speed encounter, particles on circular orbits$-$in particular those in the outer disc$-$will sweep out only a  small fraction of their orbit during Sgr's passage. Thus, we approximate the gravitational influence of Sgr as an impulsive encounter and neglect the motion of the particles in the galaxy by fixing the Galactocentric radius of each particle and Sgr, $\textit{R}_{\star}$ and $\textit{R}_{\rm sgr}$ respectively, to their positions at the time of pericenter. The change in velocity for stars from the applied impulse is obtained by integrating $a_t (t)$. In the impulsive regime, 

\begin{equation}
    \Delta \textbf{v}_{t} = \int^{\infty}_{-\infty} \textbf{a}_t (t) dt \simeq \textbf{a}_t \tau_{\rm enc}.
    \label{dv.eqn} 
\end{equation}

\begin{figure*}
	\centering
  	\includegraphics[width=1 \linewidth]{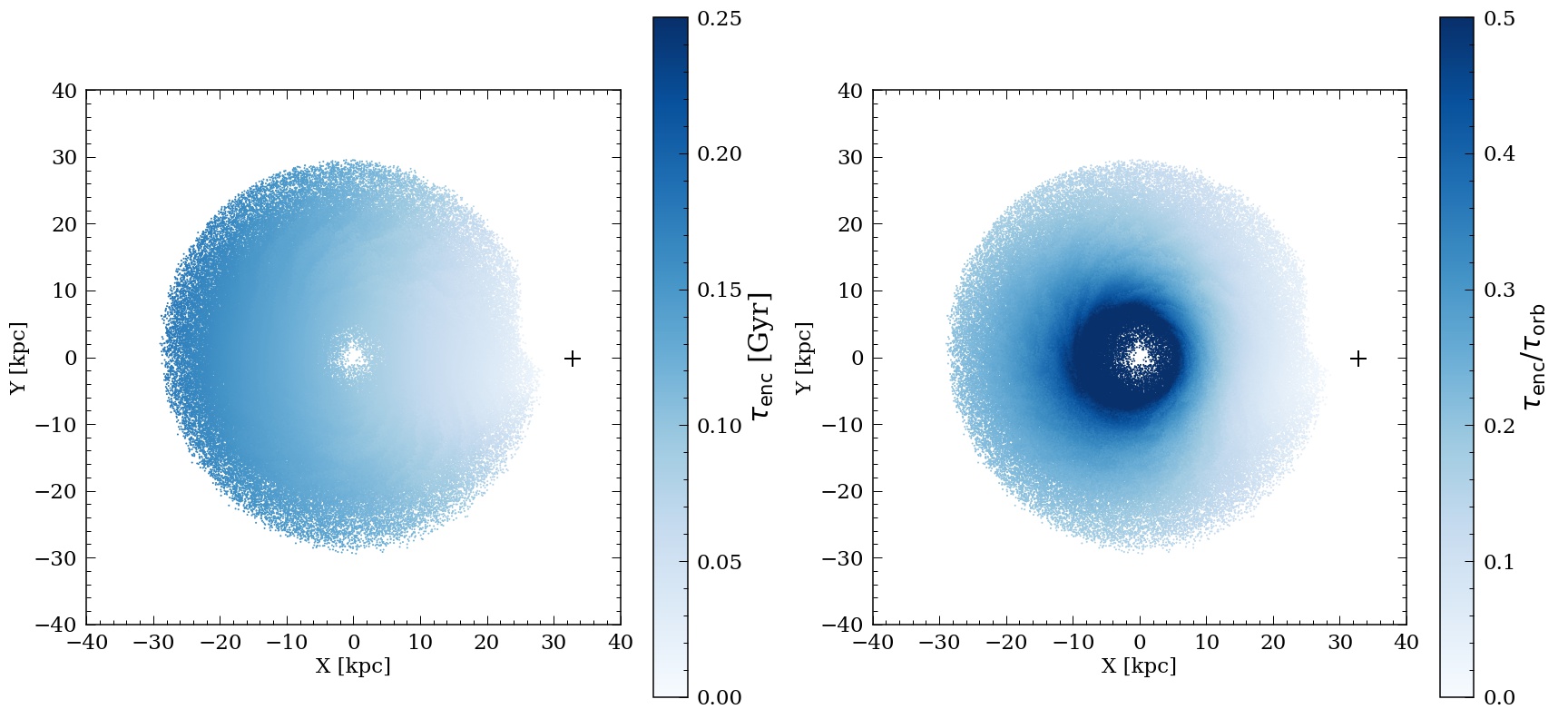}
    \caption{Estimated encounter time $\tau_{\rm enc}$ (\textbf{left}) and ratio of encounter time to disc orbital time $\tau_{\rm enc}/\tau_{\rm orb}$ (\textbf{right}) for circular orbit populations during the first Sgr encounter. The particles and Sgr (\textbf{black cross}) are positioned at their locations at the time of pericenter ($t=2.35$ Gyr).}
    \label{fig: tenc}
\end{figure*}

In Figure \ref{fig: tenc}, we estimate the encounter time for each particle to be the ratio of the distance between the particle and Sgr at pericenter, $R$, and the total velocity of Sgr, $V_{\rm sgr}$, taking the form $\tau_{\rm enc} \sim R/V_{\rm sgr}$. For the first close passage of Sgr, the average encounter time is on the order of $\tau_{\rm enc} \sim 100$ Myr. On the righthand plot of Figure \ref{fig: tenc} we compare this estimate of $\tau_{\rm enc}$ to the disc orbital time, where we use the approximate form $\tau_{\rm orb} \sim 2 \pi R_\star / v_\phi$. From this comparison, we find short encounter times for particles on the near side of the disc closest to Sgr's disc crossing, and that encounter times are longest on the far side of the disc. The ratio of $\tau_{\rm enc}$ to $\tau_{\rm orb}$ is small for outer disc particles on the near side, with the encounter time less than a tenth of the orbital time. The ratio remains quite small across the disc until reaching disc populations in the inner disc within a radius of $\sim$ 10 kpc, where $\tau_{\rm enc} / \tau_{\rm orb}$ approaches 0.5 or greater.

 \begin{figure*}
	\centering
  	\includegraphics[width=1 \linewidth]{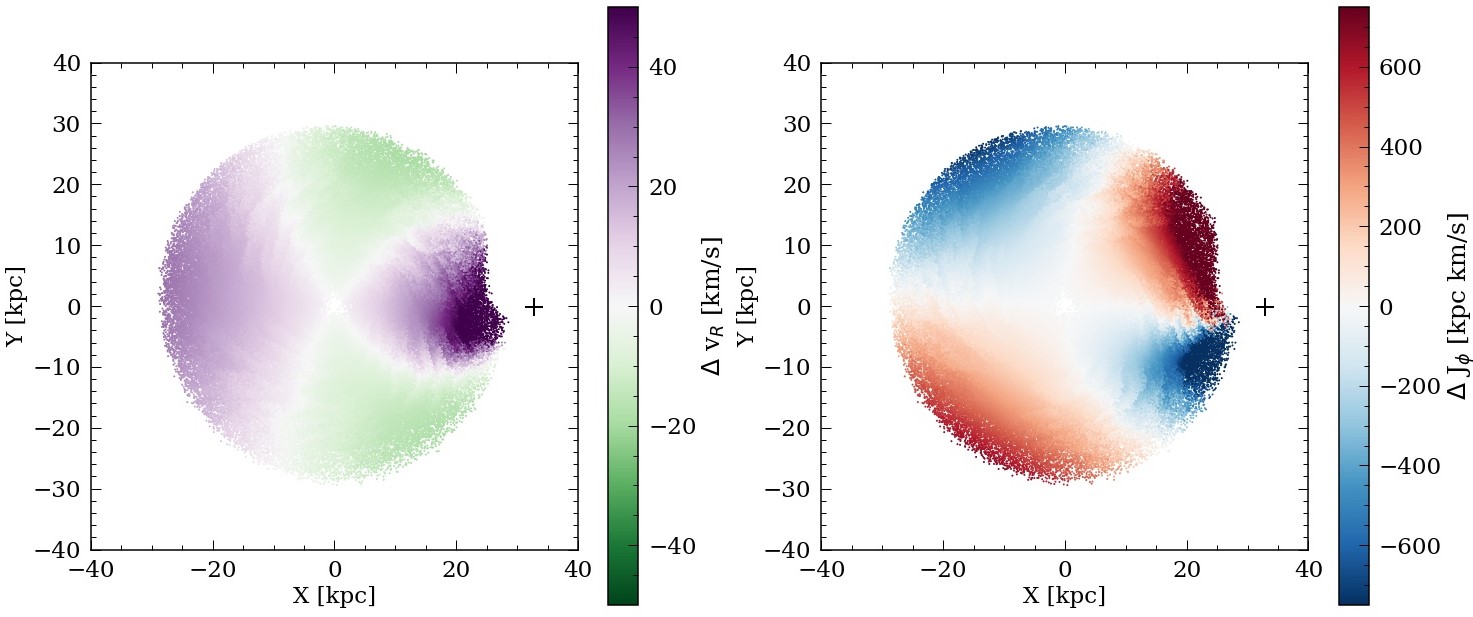}
    \caption{Analytical estimates for the changes in radial velocities (\textbf{left}) and angular momenta (\textbf{right}) for circular orbit populations using the impulse approximation for the first Sgr encounter. The particles and Sgr (\textbf{black cross}) are positioned at their locations at the time of pericenter ($t=2.35$ Gyr). The blue-red (green-purple) color gradient captures the intensity of the change estimated over the duration of the encounter.}
    \label{fig: impEST}
\end{figure*}

We test the utility of the impulse approximation by applying this simple analytic estimate to the global change in radial velocities and angular momenta during the disc's encounters with Sgr. The change in the radial velocity in the impulsive regime can be approximated from the radial component of the impulsive change in the total velocity, 
\begin{equation}
    \Delta v_R \sim \Delta \textbf{v}_{t} . \frac{\textbf{R}}{R}.
\end{equation}

Our estimate for $\Delta v_R$ informs our estimate for the impulsive change in the radial action. First, we begin with the formal definition of a general action $J_i$
\begin{align}
    J_i = \frac{1}{2 \pi} \oint_{\gamma_i} \textbf{p} d\textbf{q} ,
\end{align}
where \textbf{p,q} are a pair of canonical conjugate coordinates to \textbf{J, $\Theta$}. Borrowing from the deviation in \citet{2019_Beane_epicycle}, we use a simplified form of the radial action that follows from the epicyclic approximation that relates the radial action to the maximum radial velocity and the epicyclic frequency of an orbit. We can acquire the impulsive change in $J_R$ by including the $\Delta v_R$ from our impulsive estimate and assume $\kappa$ to be the epicyclic frequency of the orbit at the time of disc crossing,
\begin{align}
    \Delta J_R = \frac{\Delta v^2_{\rm R, max}}{2 \kappa}.
\end{align}

As for the angular momentum, the change is the cross product of the position vector of disc particle $\textbf{r}_{\star}$ and the change in velocity vector:
\begin{equation}
    \Delta \textbf{L} \sim   \textbf{r}_{\star} \times \Delta \textbf{v}_{t}.
\end{equation}
We are interested in changes to the azimuthal action or angular momentum perpendicular to the disc plane, \begin{equation}
\Delta J_\phi \equiv \Delta L_z.
\end{equation}

Figure \ref{fig: impEST} shows the change in angular momenta and radial velocities predicted from the impulse approximation, which produce a global quadrupole moment shifted in phase with respect to one another across the face of the disc. For $v_R$, there is a stark increase in $v_R$ for particles along the radial axis that extends from the Galactic center to Sgr. The sites of decreasing $v_R$ reside in the adjacent sections of the disc (north-south with respect to the Galactic center). The positive and negative values of $\Delta v_R$ map to the directions of the direct tidal forces from Sgr.  
There is a similar story for the angular momenta, where the particles that experience the greatest positive change in $J_\phi$ are those that are on orbits that place them closest to pericenter, and those on the far side of the disc exposed to a positive torque with respect to the Galactic center, which is in free-fall. Disc particles on orbits that have just passed the region of Sgr's eventual pericenter experience the greatest negative change in $J_\phi$. This is due to the presence of Sgr tugging in the opposing direction of the orbit's rotation. The tidal forces from Sgr contribute to a less, but still distinct decline in $J_\phi$ for particles that find themselves on the other side of the disc at the end of the encounter. 

 \begin{figure*}[h!]
	\centering
  	\includegraphics[width=0.9\textwidth]{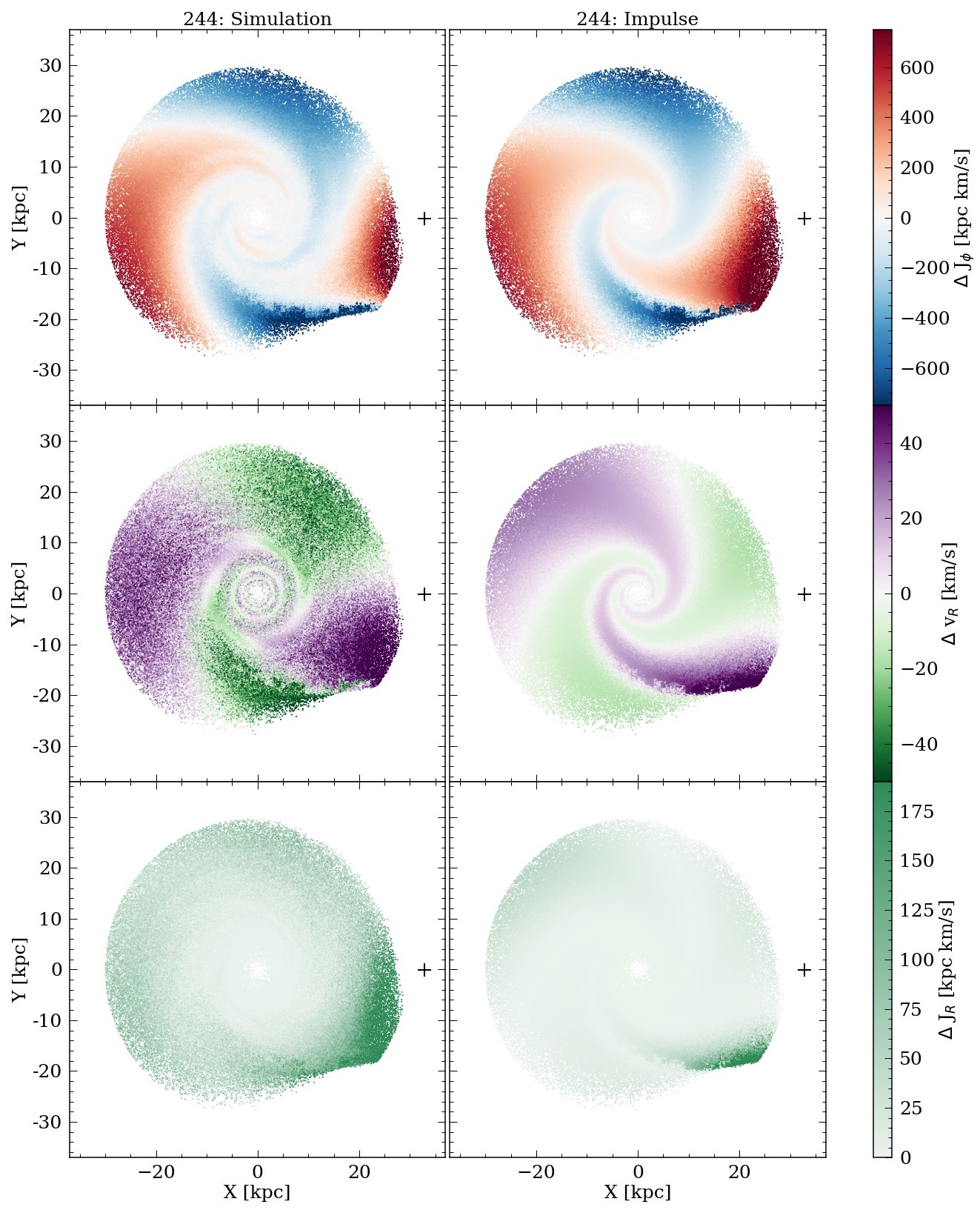}
    \caption{A comparison between the change in azimuthal action (\textbf{top row}), radial velocities (\textbf{middle row}), and radial actions (\textbf{bottom row}) for circular orbit populations observed in the simulation (\textbf{left}) and the analytical estimates (\textbf{right}) determined from the impulse approximation. The particles and Sgr (black cross) are positioned at their locations at t=2.44 Gyr, at the end of the first encounter. The blue-red (green-purple, green) color gradient captures the intensity of the change observed over the encounter. }
    \label{fig: imp244}
\end{figure*}

\subsubsection{Global Patterns of $\Delta v_R$, $\Delta J_\phi$, and $\Delta J_R$ across the Face of the Disc}
Figure \ref{fig: imp244} displays how the changes in radial velocity and angular momentum predicted from the impulse approximation are able to reasonably reproduce the global patterns of change in these quantities for circular orbits immediately following the first encounter. The timescale we chose for this comparison is based on the maximum encounter time for particles in the disc, which for the first passage is $\tau_{\rm enc} = 180$ Myr. Assuming that pericenter passage at $t=2.35$ Gyr occurs at the middle of this interval, that places the end of the encounter at $t=2.44$ Gyr. This choice of timescale to make our comparison$-$at the end of the longest $\tau_{\rm enc}-$is to ensure that we fully capture the change in dynamics transpiring during Sgr's first encounter with the disc. 

Of the three quantities observed, the impulse approximation is most successful in reproducing the global changes in angular momenta following the first encounter. This success is seen most readily in the outer disc, where the orbital times for particles is longest with respect to their encounter times. However, regions of the disc where the encounter time approaches the orbital orbital time, such as the inner $\sim 10$ kpc of the disc, the quadrupole pattern is faint, and already in the process of becoming phase-mixed. In addition, the choice to compare to the simulation at the maximum $\tau_{\rm enc}$ means that regions of the disc that have had shorter encounter times, such as the near side of the disc with respect to Sgr, may have experienced greater self-interactions and angular momentum exchange in the time before we make our comparison at $t=2.44$ Gyr. These internal disc dynamics will contribute to discrepancies between the simulation and the impulse approximation. 

As for the other quantities, the impulse approximation does an adequate, but less successful job, at matching the global patterns across the disc. The impulse approximation predicts a large change in radial velocity for a small population of particles closest to pericenter along with far milder changes for the rest of the galaxy. This produces a global structure in $v_R$ that is less symmetrical than what was observed in $J_\phi$. Some of this disagreement is because the comparison itself is inappropriate: the radial velocities will oscillate over the timescales of the interaction.

The shortcomings of using the impulse prediction for the oscillating quantity $\Delta v_R$ and comparing to disc properties at a later time should not be an issue  in our estimates for $\Delta J_R$. However, a larger population of particles in the simulation experience an enlargement in their $J_R$ following the first passage compared to the impulse prediction. These particles are concentrated mostly on the near side of the disc, consistent with the pattern seen in $\Delta v_R$. Another noticeable disagreement is the inability of the impulse approximation to appreciably capture the radial-dependence of the change in $J_R$. Particles in the outer disc across all azimuth experience some increase in $J_R$, which is something not seen to the same extent in the impulse estimates. 
.  

\begin{figure}
	\centering
  	\includegraphics[width=1\linewidth]{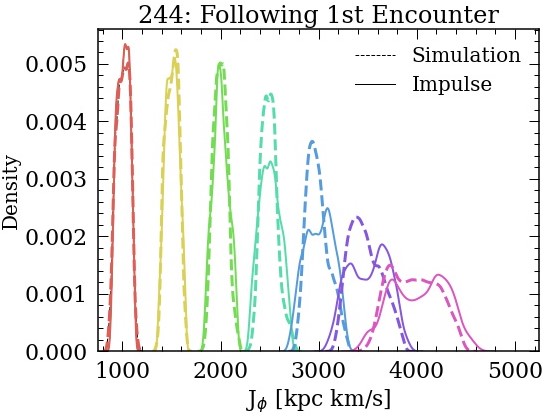}
    \caption{A comparison between the resultant $J_\phi$ and $R_g$ distributions for different mono-age/mono-$J_{\phi,0}$ populations following the first passage of Sgr at t=2.44 Gyr (\textbf{dotted}) and the predicted distributions for those same populations using the impulse approximation (\textbf{solid}).}
    \label{fig: Lz_dist}
\end{figure}

 \begin{figure*}
	\centering
  	\includegraphics[width=0.8\linewidth, scale=2]{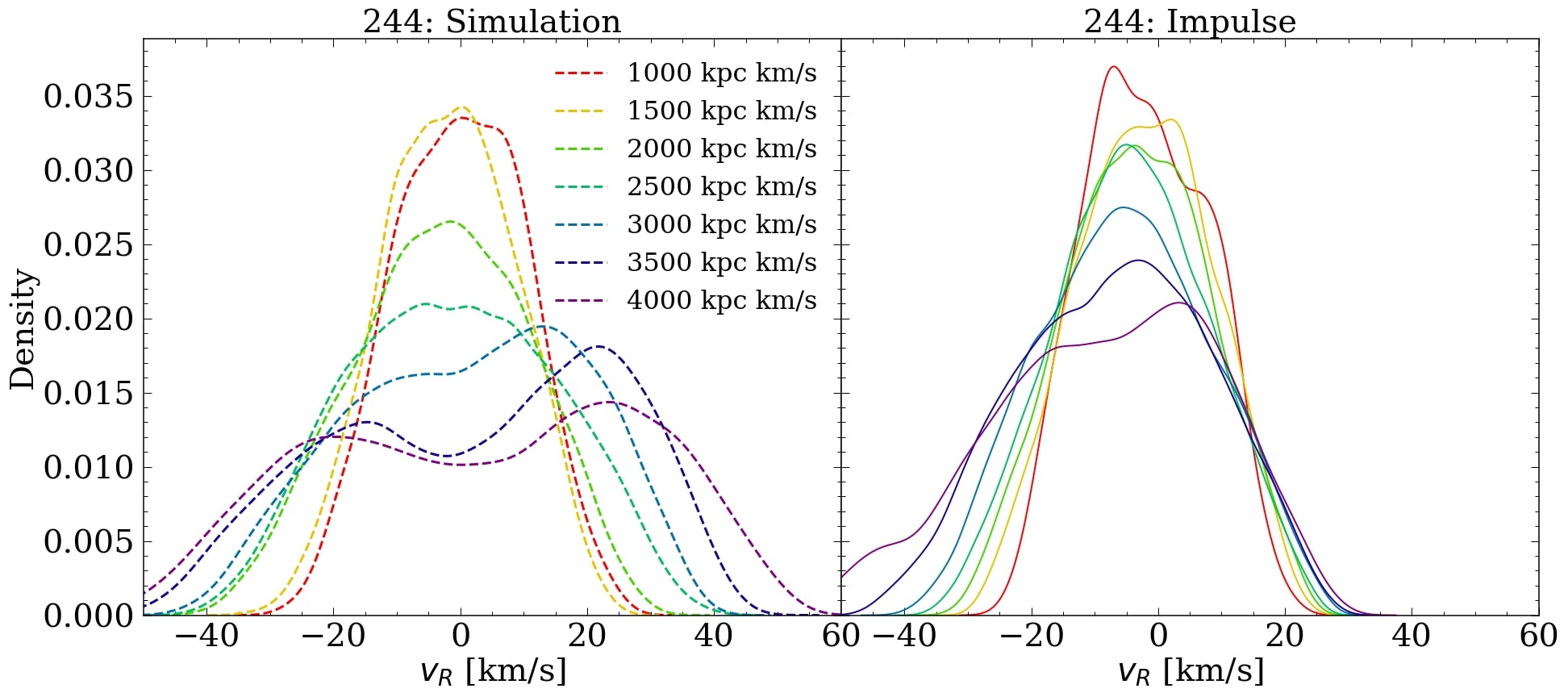}
    \caption{A comparison between the resultant $v_R$ distributions for different mono-age/mono-$J_{\phi,i}$ populations following the first passage of Sgr at t=2.44 Gyr (\textbf{dotted}) and the predicted distributions for those same populations using the Impulse approximation (\textbf{solid}).}
    \label{fig: vr_dist}
\end{figure*}

\begin{figure*}
	\centering
  	\includegraphics[width=0.90\linewidth, scale=1]{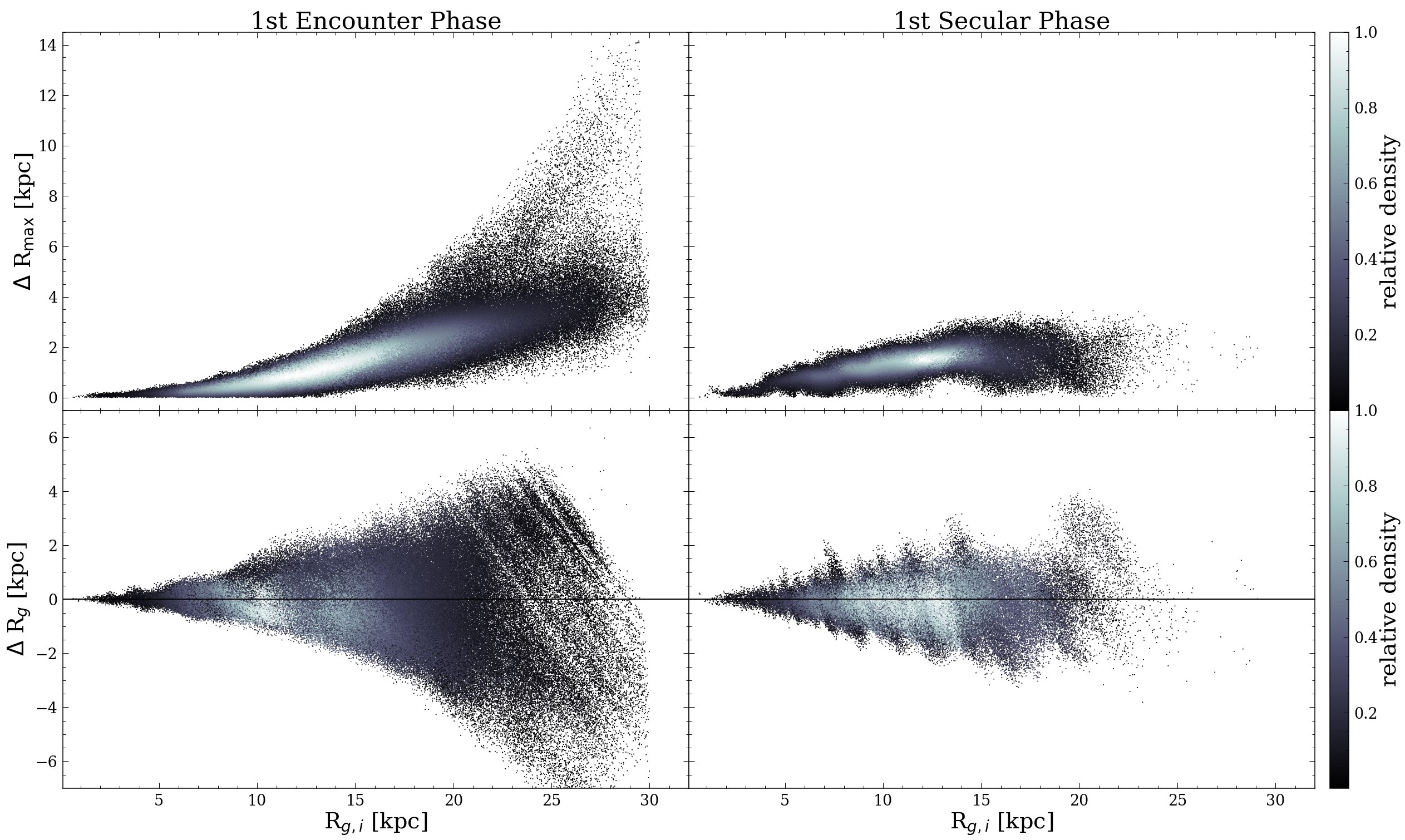}
    \caption{The maximum radial excursion, $\Delta R_{\rm max}$, and change in guiding radius, $\Delta R_{g}$, for particles on initially circular orbits plotted against their initial guiding radius $R_{g,i}$. Circular orbits were selected at $t=2.26$ Gyr at the beginning of the first crossing of Sgr (\textbf{left column}) and their $\Delta R_{\rm max}$ and $\Delta R_{g}$ were recorded at the end of the encounter at $t=2.44$ Gyr. The same is done for a selection of circular orbits at $t=3.0$ Gyr (\textbf{right column}), with their $\Delta R_{\rm max}$ and $\Delta R_{g}$ recorded at $t=4.3$ Gyr. Particles are colored by their relative density in order to better resolve structure.}
    \label{fig: imp_sec}
\end{figure*}

\subsubsection{$v_R$, and $J_\phi$-Distributions after First Close Encounter}
Here we show the distributions of $v_R$ and $J_\phi$ in the disc for different orbital populations in order to better characterize the effects of Sgr's first crossing and the impulse approximation's ability to model the scale of the disc's response. A sample of circular orbits at different annuli prior to the passage of Sgr were selected using our circular orbit criteria. We describe these particles as ``mono-age/mono-$J_{\phi,i}$ populations" due to their common orbital properties, mimicking stellar populations born at the same time on roughly circular orbits at a shared radius from the Galactic center.

Displayed in Figure \ref{fig: Lz_dist}, the colored distributions define seven distinct mono-age/mono-$J_{\phi,i}$ populations with $J_{\phi,i}$ within 200 km/s kpc bins centered around $\overline{J_{\phi,i}}=\{1000,1500, 2000, 2500, 3000, 35000, 4000\}$ kpc km/s. The dotted and solid distributions show the resulting $J_\phi$-distributions for these populations following the first crossing of Sgr and the predicted distribution of $J_\phi$ from the impulse approximation, respectively. Similiar to the azimuthal pattern of angular momentum change, the predictions from the impulse approximation are in qualitative agreement with the $J_\phi$-distributions of circular orbits in the disc following the encounter. However, it should be noted that the agreement in the spread of $J_\phi$-distributions in the inner disc, $\overline{J_{\phi,i}} \lesssim 2500$ kpc km/s, may have more to do with the fact that these inner-disc populations are largely insulated from the effects of the encounter. Their comparable orbital times with respect to the encounter timescale causes their orbits to be deformed adiabatically, so the distributions before and after Sgr should remain largely unchanged by the encounter. 

In the outer disc where Sgr is most disruptive, the impulse approximation is in nice agreement with the orbit response, reproducing the spread of each distribution. The impulsive picture diverges from the disc orbit response near the peaks of the distributions, producing distributions that appear more bimodal than normal. As discussed in the previous section, this result is likely a product from phase-mixing in the disc. 

As for the radial velocity distributions, we show in Figure \ref{fig: vr_dist} that the impulse estimates fail to capture the full bimodal behavior observed in the radial velocity distributions following the Sgr encounter. The impulse estimates underpredicts the positive change in radial velocity for particles on the near side of the disc closest to pericenter, manifesting as a strong asymmetry on the negative end of the distributions. The weak bimodality that is present in the impulse distributions can be attributed to populations on the far side of the disc during the encounter and on the adjacent sides, increasing and decreasing in $v_R$ respectively. The bimodal $v_R$ distributions in the simulation may speak to a more symmetric response in the radial velocities across the face of the disc. Alternatively, since $v_R$ is not a conserved quantity of an orbit, particles may be oscillating in the radial direction during the encounter, a contribution to the radial velocities that is not captured by the impulse approximation. 

\subsubsection{Migration and Mixing Due to the First Satellite Encounter}
The left hand panels of Figure \ref{fig: imp_sec} summarize Sgr's impact on the mixing (upper panel) and migration (lower panel). To get a better sense of the degree of mixing, an indication of radial heating, we look at the maximum radial excursion $\Delta R_{\rm max}$ after the encounter vs guiding radius $R_{g,i}$ for particles initially on circular orbits across the disc. The radial excursion of the orbit, defined as the difference between the apocenter and pericenter radius of an orbit, should begin quite small for our selection of near circular orbits, but then should be amplified during the encounter from tidal forces from Sgr. The guiding radius, $R_g$, is equivalent to the Galactocentric radius for a circular orbit with the same angular momentum, where $R_g = J_\phi / v_{\rm circ} (J_{\phi})$.  Looking at $\Delta R_{\rm max}$ and $\Delta R_g$ provide a more physically intuitive representation of mixing and migration across the disc than what can be gained by using quantities such as $\Delta v_R$, $\Delta J_R$, and $\Delta J_\phi$. 

We find a clear increasing trend between $\Delta R_{\rm max}$ and $R_{g,i}$ that is strongest for orbits in the outer disc. There is a distinct population visible within the plot that experience extreme changes in their maximum radial excursion beginning at $R_{g,i} \gtrsim 15$ kpc, reaching radial excursions as high as $14$ kpc for the most fringe disc populations. These are particles on the near side of the disc with respect to Sgr, where the disruption due to the satellite's crossing is most pronounced. On the other hand, the trend between amplifying $\Delta R_{\rm max}$ and $R_{g,i}$ is much weaker within $R_{g,i} \lesssim 13$ kpc, consistent with picture that the inner disc is largely insulated from the encounter. 

As for the change in guiding center, the $\Delta R_{g}$ for disc particles with $R_{g,i} \gtrsim 12$ kpc exhibit at least three visually distinct flares. The dense flaring occurring symmetrically around the $\Delta R_g=0$ axis captures the broad radial-dependence of the change in guiding radius prompted by the first encounter with Sgr. Significant flaring is observed for a small distinct population of particles near the edge of the disc with $R_{g,i}>18$ kpc below the x-axis, indicating a significant loss in $R_{g}$. Interestingly, this flaring signature starts almost exactly where the flaring in $\Delta R_{\rm max}$ begins during the encounter, suggesting these particles closest to Sgr's disc crossing experience strong changes in both radial excursion and guiding center. This agrees with the overall quadrupole response of the disc observed in Figure \ref{fig: imp244}, where a segment of the outer disc population experiences large changes in their radial velocities and angular momentum from strong radial forcing and torquing respectively during the first crossing.

\subsection{Secular Evolution of the Disc} \label{sec: 4.2}
Our analysis makes clear that the disruption of orbits in the outer disc is one of the key characteristics of Sgr's effects of the galaxy. This is most apparent in the radial heating of orbits as seen in $\Delta R_{\rm max}$, the broadening of $\Delta v_R$ distributions and $J_{\phi}$-distributions. However, the extended periods in the simulation between close encounters allows us also to examine the secular evolution of particle orbits in the disc. Here we distinguish ``secular" migration from Sgr-induced migration by considering the evolution of particle orbits on timescales exceeding a single rotation time, and recognize the contribution to radial migration from non-axisymmetric structures, namely the emergent spiral arms that come to prominence in this era between interactions. 

\subsubsection{Spiral Structure}
The early Sgr encounters with the disc excite the formation of transient spiral arms, corresponding to the dominance of the m=2 mode in the disc's mass density. These distortions are strongest outside the solar radius and should be the site of active $J_\phi$-exchange during the secular phase. Figure \ref{fig: density} shows that the inner disc within the solar radius largely retains its axisymmetry until the last two billions years of the simulation, where strong spiral structure and disc instability form a bar. 

The spiral structure orbits with pattern speed, $\Omega_p$, and due to the winding behavior of the spiral structure, the non-axisymmetric structure produces a pattern speed that is radius-dependent. Stars with matching angular frequency to the spiral structure are caught in a co-rotation resonance, where resonant interactions with the non-axisymmetry can alter the stars' orbital properties and $J_\phi$-evolution with no radial heating. The Lindblad resonance, a resonance between the epicyclic frequency of an orbit and the pattern speed of structure, can also drive migration, altering both $J_\phi$ and $J_R$ for resonant orbits \citep{sellwood_recent_2010}.

\subsubsection{Migration and Mixing During Secular Evolution}
Following the first crossing, the disc settles into a new equilibrium state on the order of a few orbital times. In the right column of Figure \ref{fig: imp_sec}, we track $\Delta R_g$ and $\Delta R_{\rm max}$ for stars on circular orbits a few 100 Myr following the first passage of Sgr, so we can characterize their orbital evolution deep within the secular phase. 

From $t = 3-4.3$ Gyr, we find that the prominent structure of the $\Delta R_{g}$ vs. $R_{g,i}$ are a series of ridges present at certain regions of the disc, the most visible of these structures appearing between $R_{g,i}$ $\sim$ 8 kpc and $R_{g,i}$ $\sim$ 20 kpc. These ridges are the most active sites of angular momentum-exchange and may overlap with the location of resonances. Comparable features emerge during the second secular phase between the second and fourth crossing between $t \sim 4.5-5.5$ Gyr and when observing the disc in isolation $t \sim$ 0-2 Gyr prior to Sgr's first passage. These ridges have been observed in the works of others attempting to study radial migration in isolated disc \citep{sellwood_radial_2002, roskar_radial_2012}. This gives us confidence that the angular momentum exchange observed in the disc during this time can be attributed to resonance interactions with emerging spiral arms. 

The $\Delta R_{g}$ vs. $R_{g,i}$ plots for the encounter phase and the secular phase display markedly different structure. During the encounter we see how the change in angular momentum is dominated by a radially-dependent trend, where increasing radius also increases the magnitude of change in angular momentum. Whereas during the secular phase, angular momentum redistribution is marked by the presence of ridges at resonances with a much weaker radial dependence. The overall change in angular momentum between the two eras are quite comparable out to about $R_{g,i} \sim 15$ kpc, beyond which changes from the encounter are by far the more dominant contributor. To emphasize the distinction in overall magnitude, $\Delta R_g$ measured for orbits with $R_{g,i} \gtrsim 20$ kpc during the first encounter is almost twice as great as the largest changes in $R_g$ observed for stars during the extended secular phrase with similar $R_{g,i}$ for their orbits. However, since the sample of orbits during the secular phase becomes sparse, it becomes harder to quantify exactly the difference in radial migration between the two eras in the disc outskirts. It is also important to note the difference in timescales between the two regimes displayed here. While the magnitude of $\Delta R_{g}$ is comparable between both phases across much of the disc, the encounter occurs on a timescale less than $\sim 0.2$ Gyr, whereas the first secular phase extends over an epoch of $1.3$ Gyr. 

The observed $\Delta R_{\rm max}$ at the end of the secular phase is comparable to what is observed during the encounter phase for orbits in the inner disc. Heating during the secular phase is likely attributable to resonance effects from spiral structure and pairwise interactions\footnote{In the context of the simulation, pairwise interactions refer to particle-particle scattering events, which can radially heat orbits through an exchange of random energy. In the Milky Way, a similar outcome can come about through interactions between stars and giant molecular clouds.}. However, just like with $R_g$, this agreement between the encounter and the secular phase is quickly abandoned in the outer disc, where the encounter provokes a significant radial mixing of orbits. 

\subsection{Comparing Secular and Impulsive Migration} \label{sec: 4.3}

Here we compare the evolution of orbits during the encounter phases and secular phases of disc evolution. We first compare briefly the migration and mixing of orbits as a result of the first encounter and the extended secular phase between the first and second passage, and then disentangle the $R_g$-evolution of orbits over the lifetime of the simulation to the present-day orbit, disambiguating the relative contributions from past encounters with the satellite and from secular exchanges. 

\subsubsection{Migration \& Mixing Patterns of Stellar Orbits} 

\begin{figure*}
	\centering
  	\includegraphics[width=0.8\textwidth]{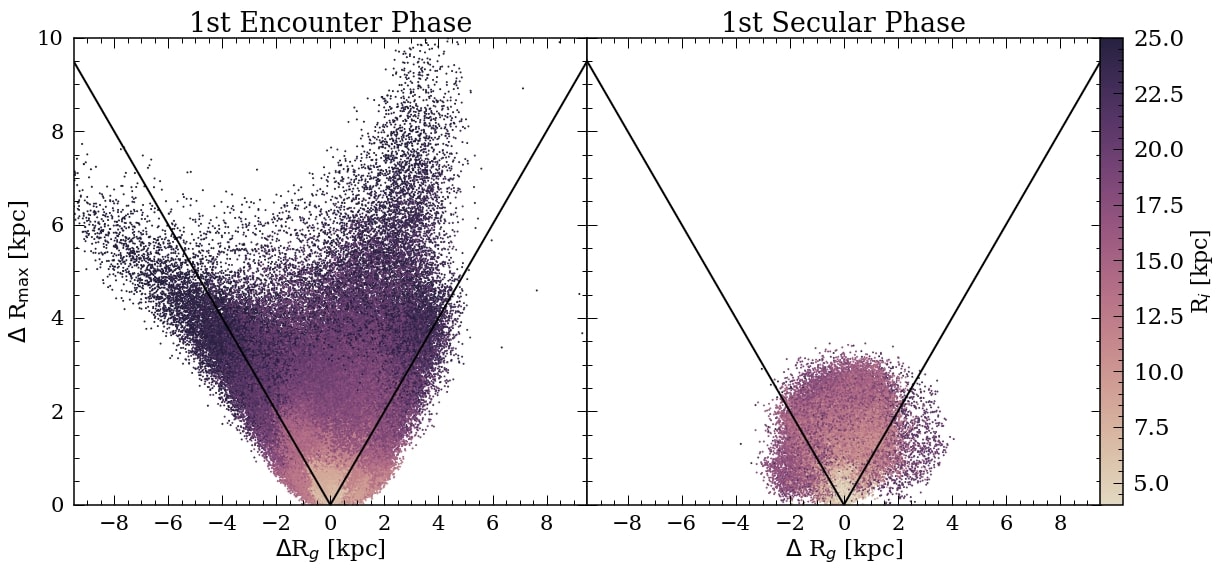}
    \label{fig: xRg_imp}
    \caption{Maximum radial excursion $\Delta R_{\rm max}$ vs. the change in guiding radius $\Delta R_g$for initially circular orbits colored by initial radius. The \textbf{left} plot displays this relation at the end of first encounter ($t=2.26-2.44$ Gyr), and the \textbf{right} plot does the same for circular orbits during the secular phase ($t=3-4.3$ Gyr). The black diagonal lines mark equality between $\Delta R_{\rm max}$ and $\Delta R_g$.}
    \label{fig: xRg_com}
\end{figure*}

The radial excursion and the change in guiding center of an initially circular orbit serve as rough proxies for the degree an orbit has increased in $J_R$ or changed in $J_\phi$ respectively. Using these quantities, we compare mixing and migration during the encounter phase and the intervening era of secular evolution. 

Figure \ref{fig: xRg_com} plots the radial excursion $\Delta R_{\rm max}$ against the change in guiding center $\Delta R_g$ for circular orbits during the first encounter and for circular orbits during the secular phase. The left-hand plot shows this relation for particles in the simulation during the encounter, revealing that Sgr imposes large charges in both $\Delta R_g$ and $\Delta R_{\rm max}$ for orbits in the outer disc. Using the diagonal line of equality as reference, the encounter radially mixes both inward and outward migrating orbits, but inward migrating stars ($\Delta R_g < 0$) experience slightly less radial heating relative to the outward migrators. This relation between mixing and migration is consistent with past work studying radial migration in a perturbed disc, which have also found that satellite encounters lead to correlated changes in angular momentum and the eccentricity of orbits \citep{quillen_radial_2009}. In addition, discs embedded in an cosmological context find a similar feature, where stars that experienced the greatest loss in circularity also experienced the strongest migration \citep{bird_radial_2012}. From the impulsive estimates shown in Figure \ref{fig: impEST}, we can gain an intuition for why we see this systematic response. Tidal forces from Sagittarius both torque and accelerate orbits most severely in the outer disc, and as seen in the left column of Figure \ref{fig: imp_sec}, both signatures are increasing functions of Galactocentric radius and most severe at azimuth closest to Sgr's passage. 

The secular phase featured in the right-hand plot of Figure \ref{fig: xRg_com} does not produce the same broad systematic effects between $\Delta R_{\rm max}$ and $\Delta R_g$ like what is observed during the encounter, nor does it produce mixing and migration to the same scale. This is not to say that there is no correlated behavior between changes in $J_\phi$ and $J_R$ during the secular era. Particular resonances, like the inner and outer Linblad resonances, can produce correlated and anti-correlated changes in these quantities \citep{sellwood_recent_2010}, but the global nature of this relation is not produced by secular modes of migration on their own. However, the relative scale of these features in the real Milky Way will depend significantly on the mass of Sgr and the strength of the spiral overdensities in the disc. 

\subsubsection{Origins of $R_g$-Distributions in Present-Day Snapshot}

\begin{figure*}
	\centering
  	\includegraphics[width=1\textwidth]{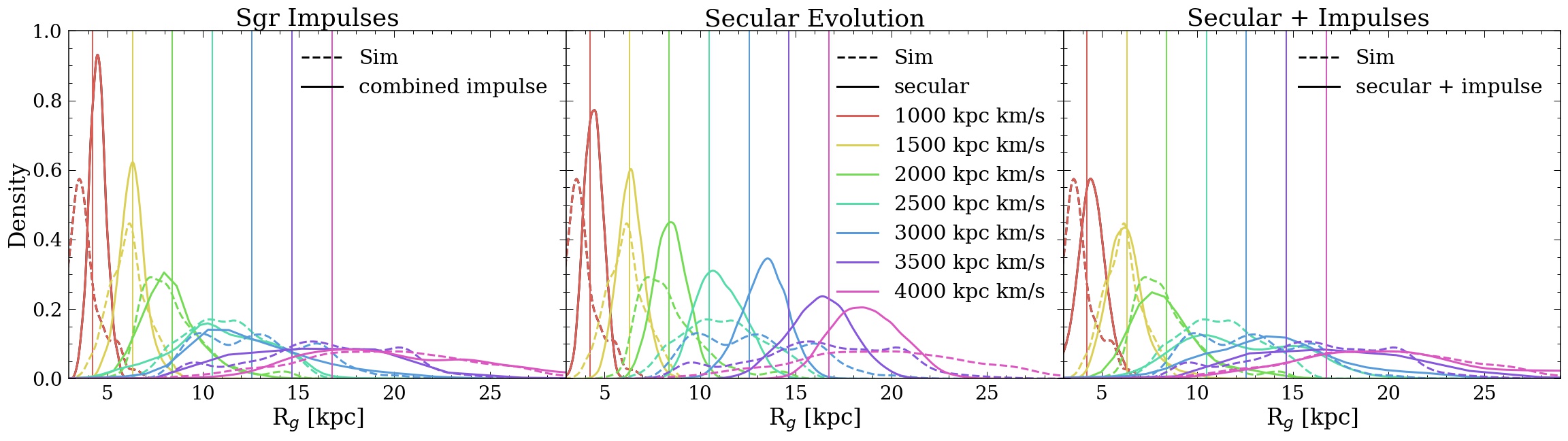}
    \caption{The dotted lines represent the resultant $R_g$-distributions at t=6.44 Gyr for particles initially selected on circular orbits at t=2.26 Gyr (\textbf{same for each panel}). The distributions are compared to the sum of the changes in $R_g$ predicted from the impulse approximation for each of the five closest pericenter passages of Sgr (\textbf{left}), the accumulative changes in $R_g$ during the two eras of secular evolution between passages (\textbf{middle}), and the $R_g$-distributions resulting from the combined contributions of the impulsive encounters and the phases of secular evolution (\textbf{right}).}
    \label{fig: comb}
\end{figure*}

After establishing the different forms of $J_\phi$-exchange during the encounter and the secular phase, here we seek to disentangle the relative contributions to the $J_\phi$-evolution of circular orbits in the simulation towards $t=6.44$ Gyr, and identify which regions of the disc each dynamical process is more influential. We again select mono-age/mono-$J_{\phi,i}$ populations at different locations in the disc at $t=2.26$ Gyr. We present their $R_g$-distributions at late times in the simulation as a more physically tractable way of gauging their overall evolution in angular momentum and migration across the disc. 

In the leftmost plot of Figure \ref{fig: comb}, we consider only the impulse-estimated contributions to the $R_g$-evolution from the five closest pericenter passages of Sgr\footnote{We ignore the third disc crossing at t$=4.83$ Gyr because Sgr crosses the midplane more than 50 kpc from the Galactic center, and therefore its impulsive contribution relative to the other pericentre passages before the present-day is negligible.}, and compare them to the $R_g$-distributions for those same orbits at $t=6.44$ Gyr. We estimate the contributions in angular momentum for each encounter using the impulse approximation as discussed in section \ref{sec: 4.1}, and present the sum of the resulting $\Delta J_{\phi}$-distribution over all five passages in terms of guiding radius, $R_g$. We find that the sum of impulsive contributions from Sgr on the disc alone reproduce the $R_g$-distributions for circular orbit populations with $R_{g,i} > 12$ kpc fairly well. This displays not only the importance of past Sgr disruptions in shaping the orbits of stars in the outer disc in the present-day, but also the power of the impulse approximation to effectively model those contributions for multiple disc crossings. 

In the middle plot of Figure \ref{fig: comb}, we compare the $R_g$-distribution for the same mono-age/mono-$J_{\phi,i}$ populations at $t=6.44$ Gyr now only to the secular contribution to the evolution, which we estimate from the sum of the changes in $R_g$ experienced during the extended periods of evolution that occur between the Sgr passages of the disc from $t=3-4.3$ Gyr, $t=4.6-5.5$ Gyr, and $t=5.65-6.05$ Gyr. The secular contribution to the $R_{g}$-distributions across the disc do not depend as strongly with increasing radius, a result that is broadly consistent with past analytic models that treat radial migration as a diffusion process in angular momentum \citep{Sanders_Binney_DFs, Frankel_cool} with no dependence on radius. However, as our work shows, these diffusion models may fail to describe the dynamics of stars in the outer galaxy where external perturbations are increasingly important, and similarly  to the case of a strong bar, may require time- and position-dependent diffusion coefficients to compare the full evolution \citep{2011A&A_brunetti_barred-diffusion}.

The right-most plot of Figure \ref{fig: comb} combines both the dynamical contributions from the impulses and secular evolution and attempts to reproduce the $R_g$-distributions at $t=6.44$ Gyr. We find that combining both the contributions from the secular phase and the impulse estimates can match with reasonable success the populations that were on circular orbits before the first crossing of Sgr with their $R_g$-distributions at late times in the simulation. Repeated passages of Sgr are the dominant contributor to angular momentum evolution in the outer disc, whereas secular effects are important across the entirety of the disc and are a prominent mechanism for migration in the inner galaxy.

\section{Implications for the Present-Day Structure of the Milky Way} \label{sec: 5}
Assuming our present-day snapshot of $t=6.44$ Gyr in our simulation, that places the most recent disc crossings with Sgr at $t=5.58$ Gyr, $t=6.1$ Gyr and $t=6.31$ Gyr, at Galactocentric distances of $12$ kpc, $9$ kpc and $24$ kpc respectively. Since not much time has passed, the disc likely has not fully equilibrated from these recent encounters. This suggests that signatures from the encounters may still be discernible in the kinematic and metallicity distribution of stars in the present-day snapshot.

Here we consider how migration driven by recent encounters with Sgr have shaped the present-day metallicity and age structure of the Milky Way. Both metallicity and ages of stars are observationally derived quantities, that can inform us on the original birth radii of stars and the overall timescale of dynamical evolution that has taken place since the present-day \citep[e.g.][]{Minchev2013, Mackereth2017,Ness2019, Ness2021, Sharma2021}.

We explore this question using our N-body simulation by painting particles with stellar populations. We assign stellar ages by isolating circular orbit populations at different timesteps in the simulation. Particles found on circular orbits at earlier times in the simulation are used to represent the dynamics of old stellar populations, while particles settling on such orbits through the course of the simulation represent populations of progressively younger ages. It is important to note that since this method to assign ages does not consider the inside-out growth of the disc surface density, this scheme should not be viewed as a faithful reconstruction of the star formation history of the Galaxy, but simply as a means to study the consequences of extended evolution of various populations over time.

Once these newly born stars are selected we assign their properties according to their location by imposing an age-dependent, negative iron metallicity gradient. Specifically, we adopt the the formalization of \citet{Frankel_cool}, and use the full derivation and fitted quantities provided therein, where metallicity is a function of age $\tau$ and birth angular momentum $J_{\phi,0}$:
\begin{align}
    \text{[Fe/H]} = \text{[Fe/H]}_{\rm max} f(\tau) + b_{\rm [Fe/H]}  +  \nabla\text{[Fe/H]} (J_{\phi,0}) \frac{J_{\phi,0}}{235 \text{ km/s}}.
\end{align}
Here the metallicity gradient, $\nabla \text{[Fe/H]} (J_{\phi,0})$, is given by
\begin{align}
\nabla \text{[Fe/H]} (J_{\phi,0}) = \left\{
        \begin{array}{llll}
          -0.03 \text{ dex/kpc} & \frac{J_{\phi,0}}{235 \text{km/s}} < 3 \text{kpc} \\
         -0.0936 \text{ dex/kpc} & \text{otherwise} 
        \end{array}
    \right. ,
\end{align}
and the constant, $b_{\rm [Fe/H]}$, which  maintains continuity between the inner 3 kpc of the galaxy and beyond is
\begin{align}
b_{\rm [Fe/H]} = \left\{
        \begin{array}{llll}
          0 & \frac{J_{\phi,0}}{235 \text{km/s}} < 3 \text{kpc} \\
         \frac{(\nabla_{\rm inner} - \nabla \text{[Fe/H]}) J_{\phi,0}}{235 \text{ km/s}} & \text{otherwise} 
        \end{array}
    \right. .
\end{align}
The time-dependence of the central metallicity, $\text{[Fe/H]}_{\rm max}$, takes the form
\begin{align}
    f(\tau) = \Big( 1 - \frac{\tau}{12 \text{Gyr}} \Big)^{\gamma_{\rm [Fe/H]}},
\end{align}
where we use the fitted value, $\gamma_{\rm [Fe/H]}=0.456$.

We note that the qualitative results of this section should not depend heavily on the particular metallicity scheme, as long as the dominant trend is a negative metallicity gradient as a function of Galactocentric radius. 

\begin{figure*}
	\centering
  	\includegraphics[width=1\linewidth]{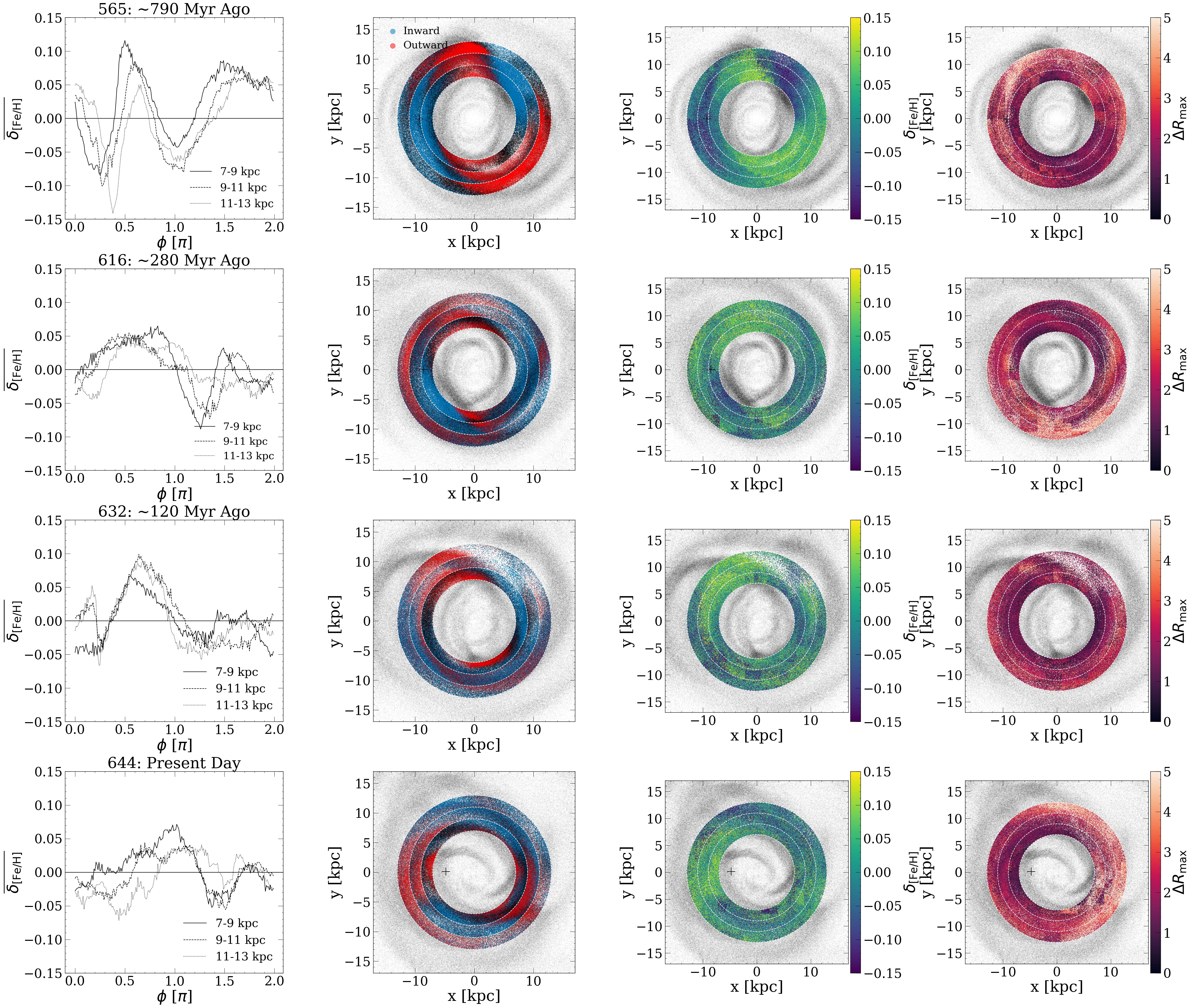}
        \caption{Azimuthally-binned average metallicity variations for stars in the annuli $7-9$ kpc, $9-11$ kpc, and $11-13$ kpc respectively (\textbf{leftmost}). Facedown maps of stars plotted in their x-y coordinates are colored by their region of origin (\textbf{left-center}), colored by azimuthal variation, $\delta_{\rm [Fe/H]}$, from the mean metallicity at their radius in the simulation snapshot (\textbf{right-center}), and colored by their maximum radial excursion (\textbf{rightmost}). In the middle column, stars colored \textbf{blue} have birth radii beyond the solar neighborhood and are inward-migrators, \textbf{red} stars are outward-migrators from the inner disc, and stars that are colored \textbf{black} are considered "in-situ", with birth radii within their respective annuli. Sgr crosses the disc at $t=5.58$ Gyr, $t=6.1$ Gyr, and $t=6.31$ Gyr before the present-day snapshot of $t=6.44$ Gyr.}
    \label{fig: azi}
\end{figure*}

\subsection{Signatures in Azimuthal Metallicity Variations from Radial Migration}

Populations that are radially migrating and mixing into the solar annulus may be distinguishable by their chemistry. If the dominant metallicity trend with radius is a negative metallicity gradient, then inward-migrating stars should be more metal-poor on average compared to in-situ populations, while outward-migrating populations with origins in the inner disc should be more metal-rich relative to the native population. If the migration induced by the most recent encounter is non-axisymmetric, similar to the quadrupole pattern in $J_{\phi}$ demonstrated in Figure \ref{fig: impEST}, then migration and mixing signatures could also manifest in the azimuthal variations of the metallicity. 

Figure \ref{fig: azi} illustrates these ideas of azimuthal variations and migration for three galactic annuli: $7-9$ kpc, $9-11$ kpc, and $11-13$ kpc. We calculate azimuthal variations by subtracting off the average metallicity of stars at a shared radius: $\delta_{\rm [Fe/H]}$ = [Fe/H](R,$\phi$) - $\overline{\text{[Fe/H]}}$(R) in radial bins of 0.2 kpc. In the top row of Figure \ref{fig: azi}, azimuthal variations from the mean metallicity are intense at $t=5.65$ Gyr, approximately $\sim 70$ Myr after Sgr's disc crossing. The disc response from a recent encounter produces correlated variations in metallicity among all three annuli. Differences in azimuthal variations of [Fe/H] can exceed $\sim 0.2$ dex between the most metal-poor and metal-rich regions of a given annulus. Inward- and outward-migrating populations densely cluster on opposite sides of the disc in a given annuli, with large and small changes to $\Delta R_{\rm max}$ dependent on their azimuthal proximity to the site of Sgr's crossing. The exact pattern of migration and mixing is contingent on the amplitude of the quadrupole-like signature in $\Delta J_{\phi}$ and excitation in $\Delta J_R$ induced by the tidal forcing from Sgr on its disc passage.  

The second row of Figure \ref{fig: azi} depicts the migration and metallicity variations a few million years following the disc crossing at $t=6.1$ Gyr. Sgr plunges into the disc at a pericentre distance of 9 kpc, and prompts an intense migration of metal-poor stars on radially heated orbits into the solar annulus from the side of the disc closest to the encounter. The response from the disc is more localized than the previous encounter, producing a weaker quadrupole signature that is mostly contained to the annuli within $7-9$ and $9-11$ kpc. 

The disc immediately following the impact at $t=6.32$ Gyr is shown in the third row of Figure \ref{fig: azi}, less than a single rotation period from the previous encounter. The pericentre passage occurs at a Galactocentric radius of 24 kpc, and triggers outward-migrating stars with large $\Delta R_{\rm max}$ to pass through all three annuli. This produces metal-rich azimuthal variations on the near side of the disc that are most intense in the outermost annuli, reaching [Fe/H] = $\pm 0.1$ dex that then decline in strength at smaller radii. The disc response from a recent encounter produces correlated variations in metallicity among annuli for all three passages, but this correlation is most readily observed at $t=5.65$ Gyr when the quadrupole-like signature is most apparent across the entirety of the disc.

The bottom row of Figure \ref{fig: azi} tracks the solar annulus to the present-day snapshot and also finds significant variations in metallicity as a function of azimuth, but slightly weaker than the azimuthal variations following the last two encounters. It is less clear that variations today can be attributed to any one passage, but encounter-induced signatures in azimuthal metallicity variations that are correlated with migrating populations remain. Average azimuthal variations reach extremum of [Fe/H] = $\pm 0.07$ dex. The maximum occurs in the solar annulus on the side of disc closest to the current position of Sgr, while the minimum is found adjacent to Sgr in the outermost annuli. In the solar annulus, inward-migrating stars with small and large $\Delta R_{\rm max}$ produce dips in metallicity azimuthal variations, with the largest of the dips corresponding to the share of the population with small $\Delta R_{\rm max}$. As the disc phase-mixes, azimuthal variations are washed out as migrated and non-migrated populations mix together and uniformly fill in the annulus. This mixing also begins to dissolve any correlation in azimuthal metallicity variations among different annuli.

\begin{figure*}
	\centering
  	\includegraphics[width=1\textwidth]{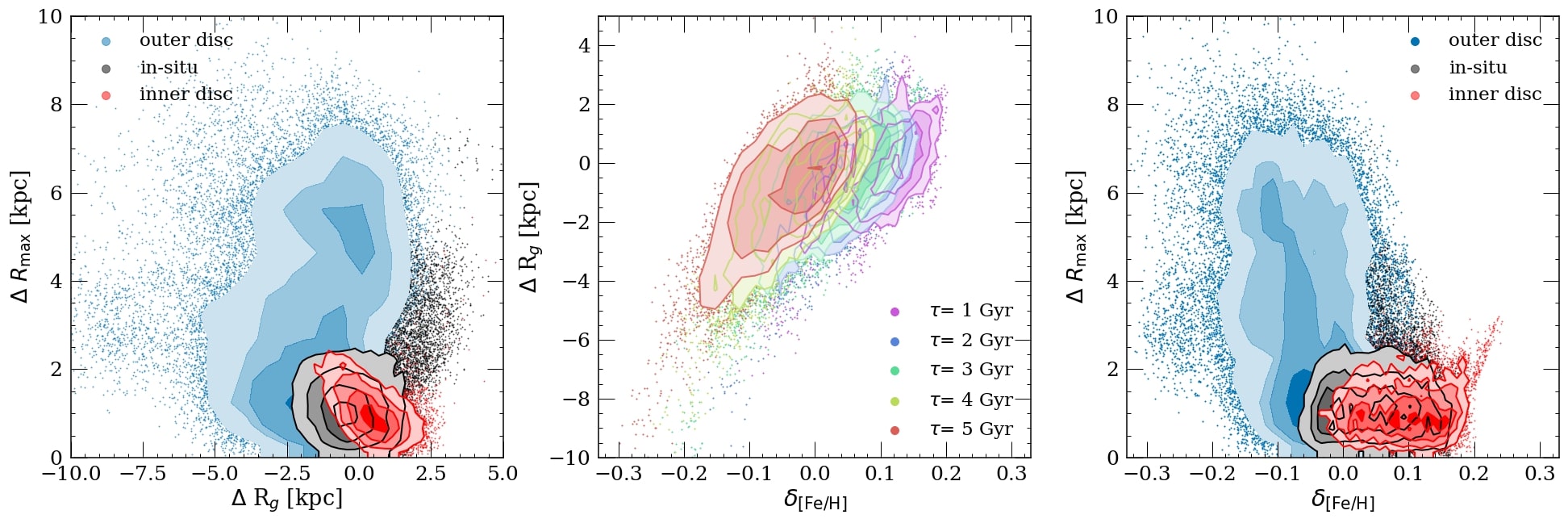}
        \caption{
        Radial excursion of present-day orbits, $\Delta R_{\rm max}$, plotted against the change in guiding radius, $\Delta R_g$, for stars of different ages within the solar annulus at the present-day snapshot of $t=6.44$ Gyr (\textbf{left}), $\Delta R_g$ vs. azimuthal variation in metallicity, $\delta_{\rm [Fe/H]}$, for those same stars (\textbf{middle}), and finally, change in guiding radius plotted against azimuthal variation in metallicity (\textbf{right}). In both the left and right plots, stars are colored by their region of origin. stars colored \textbf{blue} have birth radii beyond the solar annulus and are inward-migrators, \textbf{red} stars are outward-migrators from the inner disc, and stars that are colored \textbf{black} are considered "in-situ", with birth radii within 7-9 kpc.}
    \label{fig: xRg}
\end{figure*}

\subsection{Observational Tracers for Sgr-Induced Radial Migration and Mixing}

Disc crossings of Sgr impose a quadrupole-like signature in $\Delta J_{\phi}$ and large excitations in $\Delta J_R$ at azimuth closest to the encounter. As demonstrated in the prior section, this means for each azimuthal ring of particles prior to the disc crossing, there are particles that experience strong changes in $J_{\phi}$
and $J_R$ dependent on azimuth relative to Sgr. The residual of this quadrupole signature in angular momentum, in addition to the signature from an influx of radially heated stars seen in the observable $R_{\rm max}$, can persist for several rotational periods and manifest as a systematic quadrupole in metallicity variations. As we will show below, migrating populations can be identified most confidently by a combination of observable signatures in $\delta_{\rm [Fe/H]}$ and $\Delta R_{\rm max}$.

The left-most plot of Figure \ref{fig: xRg} displays the $\Delta R_{\rm max}$ and $\Delta R_g$ for stars in the solar annulus. We recover the correlative relation in radial excursion and migration in guiding center for stars left over from the most recent encounter. Unlike Figure \ref{fig: xRg_com} where we plot $\Delta R_{\rm max}$ vs. $\Delta R_g$ across the whole disc from the first encounter, the signature from Sgr is more localized, and the inward migration of radially heated stars is what is visible in the solar annulus from the recent passages at 9 kpc and 24 kpc. Many inward-migrating stars, colored blue in Figure \ref{fig: xRg}, have migration properties that overlap with in-situ populations and have experienced changes in $R_g$ and $R_{\rm max}$ on the scale of $\sim 2$ kpc, but there is a significant portion of them that have migrated and mixed well-beyond what is observed in the in-situ and outward-migrating populations. The cluster of stars directly above the in-situ population with $\Delta R_{\rm max} \sim 4-8$ kpc experience changes in guiding radii on the order of $0-2$ kpc, representing stars that have undergone strong mixing into the solar annulus while not fully migrating into the region. On the outer hand, almost all outward migrating stars, colored in red, are dynamically indistinguishable from the in-situ population in black. 

Reintroducing the connection between metallicity and radial migration from the prior section, we plot the relation between $\Delta R_{g}$ and $\delta_{\rm [Fe/H]}$ for stars of different ages in the solar annulus in the middle panel of Figure \ref{fig: xRg}. Just as we would expect for a disc with a negative metallicity gradient, we find that the most metal-poor stars in the annulus are the stars that have migrated the farthest inward into the galaxy, while the same can be said for metal-rich stars too from the inner galaxy. This is true for all mono-age populations, and we do not recover any clear differences in the dynamics of old and young stellar populations.

The left and middle plots of Figure \ref{fig: xRg} show that $\Delta R_g$ is related to $\Delta R_{\rm max}$ and $\delta_{\rm [Fe/H]}$ respectively, connecting the historical evolution of guiding radius to two properties that represent present-day features of a star's orbit and metallicity. We can exploit these correlations in order to project this history of radial migration into a space charted out by these present-day quantities. The right-most plot in Figure \ref{fig: xRg} swaps out $\Delta R_g$ with $\Delta R_{\rm max}$, displaying two observational quantities, the radial excursion of the orbit $R_{\rm max}$ and the azimuthal variation $\delta_{\rm [Fe/H]}$ at its present radius. We find that a portion of inward-migrating stars from the outer disc occupy a unique space in the middle and right plots of figure \ref{fig: xRg}, and that outward-migrating stars are largely degenerate with in-situ populations. A significant portion of inward-migrating stars are both highly eccentric relative to the in-situ population and quite metal-poor, approaching azimuthal variations of [Fe/H] $\sim 0.2$ dex, with respect to other stars at a common radius. This suggests that the most recent encounters with Sgr should lead to a population of metal-poor stars on eccentric orbits with orbital apocenters that cause them to mix into the solar annulus for a period during their orbits, while at the same time, other metal-poor stars that fully migrate into the region on less heated orbits that are harder to parse from more native populations.

\section{Discussion} \label{sec: 6}
\subsection{Model Limitations}
Although the properties of our simulation and the observed Milky Way-Sgr system share a broad similarity, key divergences between the model and the real galaxies place important caveats on the interpretation of our work. We have neglected other influences beyond the Milky Way and Sgr, like the LMC, because it is sufficiently distant and thought to be on its first infall, but its presence should be important to the story \citep{laporte_response_2018,laporte_influence_2018, 2021_vasiliev_tango3}. The true magnitude of the disc's response will depend on the mass-loss history of Sgr on its decaying orbit, for which our simulation is only one realization of that history. In our simulation, Sgr makes its most recent disc crossing $\sim 130$ Myr ago at $24$ kpc, and persists to the present day snapshot with a total mass of $\sim 6 \times10^{9} M_{\odot}$. Simulations such as these with more massive progenitors for Sgr with halo masses of $10^{10-11} M_{\odot}$ have had success in reproducing outer disc structure \citep{purcell_sagittarius_2011, laporte_influence_2018}. This is consistent with observations of the metallicity distribution of stars in the stream \citet{2017_gibbons_tails}, and in the core \citep{2020_Nidever_LMC}, suggesting an original mass prior to infall comparable to the LMC. Our work attempts to get a sense of the scale a more massive Sgr could have contributed to migration \& mixing in the disc from its early passages. 

The mass evolution of Sgr remains a contentious arena of debate, with early N-body models, concerned with modeling the properties of the Sgr stream, estimating remnant masses in the range $2_{-1.0}^{+1.3} \times 10^8 M_{\odot}$ \citep{Law_Majewski2010}, in rough agreement with the mass of $\sim 4 \times 10^8 M_{\odot}$ found in \citet{2020_vasiliev_sgr}. A less massive Sgr in the present day will carry significant implications for our results. Our predictions for a signature in the metallicity azimuthal variations of stars in the solar annulus and the outer disc from the most recent disc crossing should be received with a degree of caution and not viewed as a quantitative prediction for the Milky Way. The detailed mass evolution of Sgr over its several disc crossings and its velocity and Galactic distance at infall will greatly influence the overall magnitude of our results. However, the qualitative description uncovered in our analysis should be representative of the disc's response, and is broadly consistent with past studies on radial migration with an eternal perturber \citep{quillen_radial_2009, bird_radial_2012}.

As for the properties of the disc, an important difference between our simulation and the Milky Way is the lack of a bar until late times in the simulation, whereas the bar in the Milky Way is thought to be a long-lived structure \citep{2019_bovy_bar}. This disparity is likely a consequence of the simulation's initial conditions of equilibrium with a non-evolving disc, which likely stands in contrast to the formation of the Milky Way's disc, where its relation to the Gaia-Enceladus-Sausage merger is still not well understood. Since galactic bars resonantly interact with disc stars similar to spiral arms, the lack of bar in the simulation means we are likely underestimating the degree of outward radial migration from the inner disc into the solar annulus. Migration prompted by the bar has also been shown in simulations to produce azimuthal variations in metallicity that correlate with bar strength, with maxima and minima variations occurring parallel and perpendicular to the bar respectively \citep{di_matteo_signatures_2013, 2021_wheeler}. Separating azimuthal variations produced from the bar and those from the most recent disc crossings of Sgr will require more detailed work. 

Another limitation but also an avenue for future inquiry, is to consider the implications for gas dynamics and the star formation history in the disc. Enhanced star formation episodes in the solar neighborhood may coincide with pericenter passages from Sgr \citep{2020_Ruiz-Lara_sgr_SFH}. For structures in the outer disc, the different star formation histories between the Monocreos Ring and the Anticenter Stream derived from their respective chemo-dynamical properties suggest that the two structures may have been excited by different passages between Sgr and the Milky Way \citep{2020MNRAS_laporte_anticenter-stream}. Grappling with the radial flows of gas in the disc, in addition to the stars, for encounters with Sgr or for massive satellites in general, could inform subsequent chemical modelling of the Milky Way, and allow for the possibility for more quantitative predictions for its impact on the Galaxy's evolution.

\subsection{Observational Prospects}
Our results show that the disc's interaction with Sgr may leave signatures that could be uncovered in spectroscopic surveys of the real Milky Way. Our work suggests the presence of azimuthal variations in metallicity for stars in the solar annulus and the outer disc as a product of Sgr's most recent disc crossing. With surveys like APOGEE and SDSS-V's forthcoming Milky Way Mapper which have extensive coverage, azimuthal variations at the level of $<$ 0.05 dex will be readily detectable \citep{2021_sdssIV_apogee17}. From \citet{Wheeler2020}, these gradients in the Milky Way look to be on the order of $\sim$0.1, using LAMOST data, in some elements (see Figure 14 in \citet{Wheeler2020}). There are also observed azimuthal gradients in the HII gas on the Milky Way on the same order, $\sim$ 0.1 dex \citep{Wenger}. Azimuthal variations appear to be a common feature of disc galaxies, and on the order of 0.02-0.03 dex \citep{Kreckel2020}. The strength and pattern of the azimuthal variations produced will ultimately depend on the properties of the Sgr-MW encounter, and will need to be separated from internal sources of metallicity variations from bars or spiral arms \citep{di_matteo_signatures_2013, 2016_grand_spirals-patterns, 2021_wheeler, 2021_Ellers_abundance_maps}.

\section{Conclusion} \label{sec: 7}
Observations and simulation results suggest that the radial redistribution of different populations has a played an important in shaping the Milky Way's current dynamical and chemical properties. In this work, we considered the influence Sgr may have had on radial migration \& mixing in the Milky Way by studying the response of a Milky Way-like disc interacting with a satellite on a Sgr-like orbit in a collisionless N-body simulation. To isolate the effect of these encounters, we modeled the first disc crossing of Sgr using the impulse approximation, and compared the analytic estimates for the changes in $J_\phi$, $v_R$, and $J_R$ for particles on initially circular orbits$-$mimicking dynamically cold zero-age stellar populations$-$to the changes observed during the simulation. We tracked these same quantities during the secular phase that transpires between the first and second passages from $t=3.0-4.3$ Gyr. We then imposed a negative metallicity gradient on the disc to observe the lingering signatures in chemistry that this dynamical history would give rise to in the distribution of metals and other observables in the solar annulus. The major findings of our work are described below:
\begin{itemize}
    \item The impulse approximation provides a reasonable estimate for the global response in angular momentum, radial velocity, and radial action experienced for orbits in the disc during a single passage of Sgr, which roughly follow quadrupole patterns across the face of the disc. This success is most apparent for changes in angular momentum in the outermost regions of the disc where $\tau_{\rm enc}<\tau_{\rm orb}$.
    \item Sgr's influence on disc orbital properties is strongest in the outer disc. This is observed in the broadening of $J_\phi$ distributions, the radial heating in $J_R$, and the bimodal behavior of $v_R$ distributions for populations with $J_{\phi,i} > 3000$ kpc km/s ($R_{g,i}\sim$12 kpc). The combined impulses from Sgr are the dominant contributor to the angular momentum evolution of stars in the outer disc. Between Sgr passages, the disc evolution is governed by mild secular evolution. Ridges are visible in the $\Delta J_\phi$ for circular orbits evolving during this time, which may overlap with the location of resonances of transient spiral modes. 
    \item The encounter with Sgr produces a systematic response in radial heating and angular momentum. This relationship is mostly readily seen in plotting the maximum radial excursion of an orbit against its change in guiding radius, a result consistent with past work on orbit responses to satellite bombardment. This strong correlation is not visible across the disc during the secular phase.  
    \item The quadrupole-like pattern in migration and mixing brought forth from the most recent encounters with Sgr manifest as strong correlated azimuthal variations in metallicity for the solar annulus and the outer disc immediately following the impact, but this feature weakens towards the present-day as the disc begins to phase-mix. A distinct signature of Sgr-induced migration \& mixing in the solar annulus is the relocation of metal-poor stars on highly eccentric orbits into the inner galaxy with a correlated structure in $\delta_{\rm [Fe/H]}$ and $\Delta R_{\rm max}$.
\end{itemize}

Overall, we conclude that discussions of radial migration and mixing in galaxies cannot be limited to secular processes alone. Both the cosmological context of hierarchical structure formation and observations of ongoing and past merger events demand that encounter-driven migration also be considered. Metallicity distributions combined with dynamics offer us a promising way of isolating these external influences. Spectroscopic surveys with an expansive azimuthal coverage, in addition to detailed kinematics from the GAIA satellite, will soon arm us with the necessary data to undergo the great task of untangling these mechanisms in our own Galaxy.

\section{Acknowledgements}
We would like to thank the Milky Way Stars group at Columbia University for great conversations and advice over the course of this project. We would also like to thank Douglas Filho for his action calculation code for orbits in the L2 model, and Suroor Seher Gandhi for contributing valuable figures. CL acknowledges funding from the European Research Council (ERC) under the European Union's Horizon 2020 research and innovation programme (grant agreement No. 852839). KVJ was supported by NSF grant AST-1715582. MKN is supported in part by a Sloan Foundation Fellowship.

\section*{Data Availability}
No new data were generated or analysed in support of this research.



\bibliographystyle{mnras}
\bibliography{report} 

\bsp	
\label{lastpage}
\end{document}